\documentclass{article}
\PassOptionsToPackage{numbers,sort&compress}{natbib}

\usepackage[main, preprint]{neurips_2026}

\usepackage[utf8]{inputenc} 
\usepackage[T1]{fontenc}    
\usepackage{hyperref}       
\usepackage{url}            
\usepackage{booktabs}       
\usepackage{amsfonts}       
\usepackage{nicefrac}       
\usepackage{microtype}      
\usepackage{xcolor}         
\usepackage{makecell}

\usepackage{amsmath}        
\usepackage{graphicx}       
\usepackage{multirow}       

\newcommand{\rankci}[3]{\makecell{#1\\{\small[#2, #3]}}}

\title{Platonic Representations in the Human Brain: Unsupervised Recovery of Universal Geometry}

\author{
  Pablo Marcos-Manchón$^{1,2}$ \quad 
  Rishi Jha$^{3}$ \quad
  Lluís Fuentemilla$^{1,2,4}$ \\
  \\
  $^{1}$Department of Cognition, Development and Educational Psychology, University of Barcelona\\
  $^{2}$Institute of Neurosciences, University of Barcelona\\
  $^{3}$Department of Computer Science, Cornell University\\
  $^{4}$Bellvitge Institute for Biomedical Research, Spain
}

\begin{document}
\maketitle

\begin{abstract}
The Strong Platonic Representation Hypothesis suggests that representational convergence in artificial neural networks can be harnessed constructively: embeddings can be translated across models through a universal latent space without paired data. We ask whether an analogous geometry can be recovered across human brains. Using fMRI data from the Natural Scenes Dataset, we propose a self-supervised encoder that learns subject-specific embeddings from brain data alone by exploiting repeated stimulus presentations. We show that these independently learned spaces can be translated across subjects using unsupervised orthogonal rotations, without paired cross-subject samples or intermediate model representations. Synchronizing pairwise rotations into a single shared latent space further improves cross-subject retrieval. These results provide evidence for a shared geometry in learned representations of human visual responses: subject-specific fMRI embeddings can be aligned across individuals within a common coordinate system.
\end{abstract}

\vspace{0.8cm} 

\section{Introduction}

The \emph{Platonic Representation Hypothesis} posits that independently trained artificial neural networks converge toward geometrically similar representations by recovering shared latent structure in the world~\cite{mikolov2013,Raghu2017,Kornblith2019,sucholutsky2025,Huh2024}. Recent constructive work pushes this hypothesis further: if representations share a common geometry, then embeddings from one model should be translatable into another model's latent space without shared inputs or paired supervision~\cite{Rishi2025,Dar2026}. However, evidence for this strong form of representational convergence has so far come almost entirely from artificial vision and language systems~\cite{Huh2024,bansal2021revisiting,lu2025survey}. Whether the same principle extends to biological neural systems remains unknown.

In neuroscience, a long line of work on inter-subject synchrony and representational similarity shows that neural responses, and the geometries they induce, can be preserved across individuals during shared stimulus processing~\cite{Hasson2004,Nastase2019,Kriegeskorte2008,Lin2024}. However, existing evidence typically depends on shared stimuli, paired measurements, or external reference spaces. Functional alignment methods for fMRI, such as \textit{hyperalignment}, map neural responses into common spaces by using shared stimuli to establish cross-subject correspondences~\cite{Haxby2011,Guntupalli2016}. Other approaches introduce anchors through model-derived feature spaces~\cite{zhou2024clipmused} or image-to-fMRI encoders~\cite{Wasserman2026}. What remains open is whether shared neural geometry can be recovered from independently learned brain representations alone.

In this paper, we extend the Strong Platonic Representation Hypothesis~\cite{Rishi2025} to the human visual cortex. We ask whether subject-specific fMRI embedding spaces, learned independently from neural data, can be translated across subjects using only the intrinsic geometry of neural responses. We evaluate this setting on the Natural Scenes Dataset (NSD)~\cite{Allen2022}, a canonical fMRI dataset of subjects viewing complex natural images. Our contributions are:
\begin{enumerate}
    \item We introduce a self-supervised encoder that learns subject-specific fMRI embeddings from repeated stimulus presentations.
    \item We show that independently learned subject embeddings can be translated across subjects using unsupervised orthogonal mappings, recovering accurate instance-level correspondences.
    \item We synchronize pairwise rotations into a single shared latent space, improving cross-subject retrieval and showing that independently learned subject spaces are mutually compatible with a common coordinate system.
\end{enumerate}

Our results provide evidence that independently learned fMRI embeddings share a geometric structure that can be recovered without paired cross-subject data, enabling neural models to transfer across individuals.

\section{Problem Setup}

Let $s \in \{1,\dots,S\}$ represent a subject who provides fMRI responses to visual stimuli drawn from a distribution $\mathcal{D}$. $X^{(s)} \in \mathbb{R}^{n_s \times v_s}$ denotes the subject's neural activity matrix, where rows correspond to image presentations and columns to subject-specific voxel responses. Our goal is to test whether these independently observed fMRI responses can be mapped into a shared latent space $\mathcal{Z}$ using only their intrinsic representational geometry, without paired cross-subject data for learning the mappings.

Subjects observe disjoint image sets independently sampled from $\mathcal{D}$. No shared images or paired cross-subject correspondences are available for learning the translations. A held-out set of shared images, observed by all subjects, is used only for evaluation.

\textbf{Subject-specific embeddings.}
For each subject, we learn a mapping $f_s : X^{(s)} \rightarrow Z^{(s)}$, with $Z^{(s)} \in \mathbb{R}^{n_s \times d}$, that projects voxel responses into a low-dimensional subject-specific space $\mathcal{Z}^{(s)} \subset \mathbb{R}^d$. The mapping $f_s$ is self-supervised from each subject's neural activity alone, without external model features or cross-subject supervision. Thus, $Z^{(s)}$ is intended to reflect the intrinsic organization of the stimulus responses of subject $s$.

\textbf{Unpaired brain-to-brain translation.}
Given independently learned embeddings $\{Z^{(s)}\}_{s=1}^S$, we seek transformations that translate each subject space into a common latent space. Specifically, we learn one transformation $R_s : \mathcal{Z}^{(s)} \rightarrow \mathcal{Z}$ per subject, such that embeddings evoked by the same image map to consistent coordinates, i.e., $Z^{(s)}R_s \approx Z^{(t)}R_t$ for subjects $s$ and $t$ on held-out shared images. Since no such paired images are available during training, the transformations must be inferred from the global geometry of the unpaired subject-specific spaces. We evaluate translation quality using cross-subject retrieval on the held-out shared images, following prior unsupervised mapping protocols~\cite{Rishi2025,Dar2026}.

\textbf{Shared geometry hypothesis.}
We hypothesize that subject-specific spaces $\mathcal{Z}^{(s)}$ are noisy instances of a shared latent geometry. Under this hypothesis, inter-subject differences should be captured by approximately isometric transformations. We therefore restrict transformations to orthogonal maps, $R_s \in \mathcal{O}(d)$, which preserve distances and inner products. This constraint prevents arbitrary geometric warping and provides a strong test of whether independently learned neural representations can be translated into a shared coordinate system using geometry alone.

\begin{figure}
    \centering
    \includegraphics[width=\linewidth]{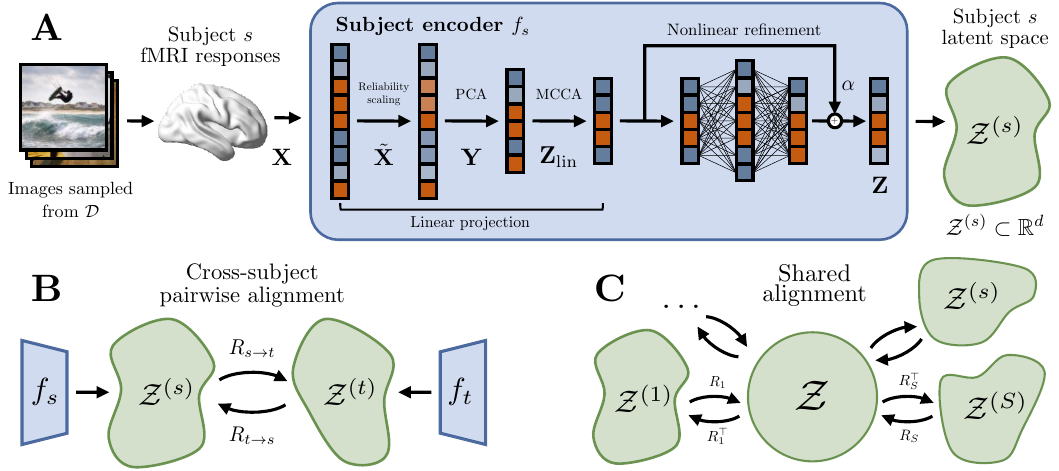}
    \caption{\textbf{Method overview.}
    \textbf{(A) Subject encoder.} For each subject, fMRI responses are mapped into a low-dimensional embedding space using voxel reliability weighting, PCA, and multi-view CCA (MCCA), followed by a residual nonlinear refinement trained from repeated stimulus presentations.
    \textbf{(B) Pairwise brain-to-brain translation.} Independently learned subject embeddings are translated between subject pairs by estimating orthogonal rotations $R_{s\rightarrow t}$ from geometry-derived pseudo-correspondences.
    \textbf{(C) Shared latent space.} Pairwise rotations are synchronized to recover one orthogonal transformation $R_s$ per subject, mapping all subject embeddings into a common space.}
    \label{fig:overview}
\end{figure}

\section{Method}\label{sec:methods}

We proceed in three stages. First, for each subject, we learn a subject-specific encoder that maps fMRI responses into a lower-dimensional embedding space. Second, we translate embeddings between subject pairs by estimating unsupervised orthogonal rotations from their geometry. Finally, we synchronize the pairwise rotations to recover one transformation per subject, mapping all embeddings into a shared latent space. An overview is shown in Fig.~\ref{fig:overview}.

\subsection{Learning repetition-invariant embeddings}

For each subject $s$, we learn a mapping $f_s:\mathbb{R}^{v_s}\to \mathbb{R}^{d}$ using repeated stimulus presentations as self-supervision. Each image is presented $r$ times, yielding multiple measurements, or views, of the same underlying neural signal. For clarity, we drop the subject index $(s)$ in this subsection. Let $\{X_i\}_{i=1}^r$, with $X_i \in \mathbb{R}^{n_s \times v_s}$, denote view-specific response matrices for the same stimuli, and let $X \in \mathbb{R}^{r n_s \times v_s}$ denote their concatenation. To mitigate temporal drifts, repetitions are randomly assigned to views for each stimulus.

Traditional fMRI denoising averages repeated measurements under independence assumptions~\cite{Liu2016, GLMSingle2022, Ozcelik2023, scotti2023reconstructing}. In contrast, we treat repetitions as noisy views of a stable latent representation and optimize $f_s$ for repetition invariance, such that for responses $x_i$ and $x_j$ to the same image, $f_s(x_i) \approx f_s(x_j)$.

\textbf{Voxel reliability weighting.}
We first reweight voxels by their reliability across repetitions. For each voxel $v \in \{1,\dots,v_s\}$, we compute reliability as the average correlation across repetition pairs:
\begin{equation}
\gamma_v = \frac{2}{r(r-1)} 
\sum_{1 \le i < j \le r}
\rho\big(X_i[:, v], \, X_j[:, v]\big).
\end{equation}

We then scale each voxel by its reliability, $\tilde{X}_i[:, v] = \max (0, \gamma_v) \cdot X_i[:, v]$, reducing the influence of voxels without stable repetition structure.

\textbf{Low-dimensional linear projection.}
We project reliability-weighted responses $\tilde{X}_i$ into a lower-dimensional subspace using PCA, obtaining $Y_i \in \mathbb{R}^{n_s \times d_{\text{PCA}}}$ with $d_{\text{PCA}} \ll v_s$. Let $Y \in \mathbb{R}^{r n_s \times d_{\text{PCA}}}$ denote the concatenation of all views.

To extract components shared across repetitions, we apply multi-view canonical correlation analysis (MCCA) to $\{Y_i\}_{i=1}^r$~\cite{MCCA1971, RegMCCA2011}. MCCA learns projections $\{U_i\}_{i=1}^r$, with $U_i \in \mathbb{R}^{d_{\text{PCA}} \times d}$, that maximize cross-view correlation:

\begin{equation}
\max_{\{U_i\}} \sum_{i < j} \rho\big(Y_i U_i,\; Y_j U_j\big).
\end{equation}

To obtain a single target representation, we project each sample in $Y$ through all view-specific mappings and average the projections:

\begin{equation}
\bar Z_{\mathrm{lin}} = \frac{1}{r} \sum_{i=1}^r Y U_i .
\end{equation}

We then regularize this mapping using ridge regression:
\begin{equation}
W^* = \arg\min_{W} \; \|Y W - \bar Z_{\mathrm{lin}}\|_F^2 + \lambda_{\mathrm{reg}} \|W\|_F^2,
\end{equation}
yielding $Z_{\mathrm{lin}} = Y W^* \in \mathbb{R}^{r n_s \times d}$.

\textbf{Nonlinear residual refinement.}
We refine the linear embedding with a residual nonlinearity:
\begin{equation}
Z = f_s(X) = Z_{\mathrm{lin}} + \alpha \, g_\theta(Z_{\mathrm{lin}}),
\end{equation}
where $g_\theta$ is a multi-layer perceptron (MLP) and $\alpha$ is a learnable scalar. We freeze the linear projection and optimize only $\theta$ and $\alpha$. Given view embeddings $\{Z_i\}_{i=1}^r$, where $Z_i=f_s(X_i)$, we use a contrastive InfoNCE loss over all view pairs:

\begin{equation}
\mathcal{L}_{\mathrm{NCE}} = \frac{1}{r(r-1)} \sum_{\substack{i,j=1 \\ i \neq j}}^{r} \mathcal{L}_{\mathrm{InfoNCE}}(Z_i, Z_j),
\end{equation}
with in-batch negatives and cosine similarity~\cite{infoNCE2019}. We also add a cosine pull term,

\begin{equation}
\mathcal{L}_{\mathrm{pull}} = \frac{2}{r(r-1)} \sum_{1 \le i < j \le r}
\big(1 - \mathrm{sim}(Z_i, Z_j)\big),
\end{equation}
where $\mathrm{sim}(Z_i, Z_j)$ is the mean cosine similarity between corresponding samples. The refinement minimizes $(\theta^*, \alpha^*) = \arg\min_{\theta,\alpha} \mathcal{L}_{\mathrm{NCE}} + \lambda_{\mathrm{pull}}\mathcal{L}_{\mathrm{pull}}$.

\subsection{Pairwise brain-to-brain translation}\label{subsec:pairwise}

To translate embeddings between two subjects, we apply \textit{mini-vec2vec}~\cite{Dar2026} to neural embeddings. Given subjects $s$ and $t$, we learn an orthogonal transformation $R_{s \rightarrow t} \in \mathcal{O}(d)$ such that $Z^{(s)}R_{s \rightarrow t} \approx Z^{(t)}$, without paired cross-subject samples during training.

We average repetitions in embedding space to obtain one representation per image, $Z^{(s)} \in \mathbb{R}^{n_s \times d}$ and $Z^{(t)} \in \mathbb{R}^{n_t \times d}$, reducing trial-level measurement error. We construct pseudo-matched pairs by clustering each space with K-means and matching centroids through their pairwise similarity structure using a quadratic assignment solver. Each embedding in $Z^{(s)}$ is then matched to the average of its nearest neighbors in $Z^{(t)}$ based on relative similarity to these matched anchors, yielding pseudo-parallel pairs $(z^{(s)}, \tilde z^{(t)})$. These pairs define the initial orthogonal Procrustes problem:

\begin{equation}
R_{s \rightarrow t}^{(0)} = \arg\min_{R \in \mathcal{O}(d)} \| Z^{(s)} R - \tilde Z^{(t)} \|_F^2.
\end{equation}

We refine the translation iteratively using an approach similar to Iterative Closest Point~\cite{ICP}. At iteration $k$, a subset of source embeddings is transformed by $R_{s \rightarrow t}^{(k)}$ and matched to nearest neighbors in the target space. These updated pseudo-targets define a new Procrustes solution $\widehat R_{s \rightarrow t}^{(k)}$, and the transformation is updated as

\begin{equation}
\tilde R_{s \rightarrow t}^{(k+1)} = (1-\beta)R_{s \rightarrow t}^{(k)} + \beta \widehat R_{s \rightarrow t}^{(k)}.
\end{equation}

After each update, $\tilde R_{s \rightarrow t}^{(k+1)}$ is projected onto $\mathcal{O}(d)$ by SVD to obtain $R_{s \rightarrow t}^{(k+1)}$.

Because the learned mappings are sensitive to clustering
initialization, we generate multiple candidate rotations per
directed subject pair using different random seeds.
We jointly select one candidate per pair by maximizing
cycle consistency across subjects: composing the mappings
from $s$ to $t$ and from $t$ to $u$ should approximate
the direct mapping from $s$ to $u$.
Selection is fully unsupervised and uses only the candidate
rotation matrices, without accessing image correspondences
or evaluation scores.
Appendix~\ref{sec:unsupervised-selection} provides the
selection algorithm and its stability analysis.

Finally, we symmetrize the selected pairwise translations
to enforce $R_{t \rightarrow s} = R_{s \rightarrow t}^\top$:

\begin{equation}
R_{s \rightarrow t} = \mathrm{Proj}_{\mathcal{O}(d)}\!\left( \frac{R_{s \rightarrow t} + R_{t \rightarrow s}^\top}{2} \right).
\end{equation}

\subsection{Shared latent space construction}\label{subsec:global}

Given pairwise translations $\{R_{s \rightarrow t}\}_{s,t=1}^S$, we construct a shared latent space $\mathcal{Z}$ by solving an orthogonal synchronization problem over $\mathcal{O}(d)$~\cite{Singer2011,WangRotation2013}. The goal is to recover one transformation $R_s \in \mathcal{O}(d)$ per subject such that $R_{s \rightarrow t} \approx R_s R_t^\top$, placing all subjects in a common coordinate system, i.e., $Z^{(s)}R_s \approx Z^{(t)}R_t$.

We form a block matrix $B \in \mathbb{R}^{Sd \times Sd}$ whose $(s,t)$-th block is $R_{s \rightarrow t}$ for $s \neq t$ and the identity for $s=t$. In the ideal noise-free case, this matrix factorizes as
\begin{equation}
B = 
\begin{bmatrix}
R_1 \\
R_2 \\
\vdots \\
R_S
\end{bmatrix}
\begin{bmatrix}
R_1 \\
R_2 \\
\vdots \\
R_S
\end{bmatrix}^\top.
\end{equation}

We recover a relaxed solution using a spectral method for orthogonal synchronization~\cite{Singer2011}. We compute the top-$d$ eigenvectors of $B$, yielding $U \in \mathbb{R}^{Sd \times d}$, whose blocks $\{U_s\}_{s=1}^S$ approximate the subject-specific transformations. Each block $U_s$ is projected onto $\mathcal{O}(d)$ by taking its closest orthogonal matrix.

The resulting transformations define a shared latent space in which each subject embedding is mapped as $Z_{\mathrm{shared}}^{(s)} = Z^{(s)}R_s$. Global synchronization denoises pairwise estimates by enforcing cycle consistency across the subject graph.

\section{Experiments}

We evaluate our method on the Natural Scenes Dataset (NSD)~\cite{Allen2022}, an fMRI dataset of 8 participants viewing natural images from COCO~\cite{Lin2014MicrosoftCC}. Each participant viewed up to 10,000 distinct images, each repeated up to three times. NSD includes subject-specific images, unique to each participant, and a smaller set of images shared across participants. We use only subject-specific, non-shared images to learn subject encoders and unsupervised brain-to-brain translations. Evaluation uses 515 held-out shared images with three repetitions for every subject. Supervised brain-to-brain baselines use additional shared images for alignment training, excluding these 515 evaluation images. See Appendix~\ref{sec:fmri-processing} for preprocessing details.

We evaluate retrieval in two settings. Within subjects, embeddings from one repetition retrieve the matching image from another repetition among 515 candidates, averaged across repetition pairs. Across subjects, repetitions are first averaged, and translated embeddings from subject $s$ retrieve the matching image from subject $t$ among the same 515 held-out images. In both settings, we report Mean Rank, R@1, and RSA. Mean Rank is the average rank of the correct image (chance $=258$, optimum $=1$); R@1 is nearest-neighbor accuracy (chance $=1/515 \approx 0.002$); RSA is the Pearson correlation between representational dissimilarity matrices.

\begin{table}[t]
\centering
\caption{\textbf{Per-subject encoder performance.}
Within-subject retrieval across repeated presentations, averaged across repetition pairs. Embeddings from one repetition are used to retrieve the matching image from another among 515 candidates. RSA is the Pearson correlation between RDMs across repetitions. Lower is better for Mean Rank (chance $=258$); higher is better for R@1 (chance $=0.002$) and RSA.}
\label{tab:within_subject_per_subject}
    \begin{tabular}{ll|cccccccc|c}
\toprule
\textbf{Method} & \textbf{Metric} & \textbf{S1} & \textbf{S2} & \textbf{S3} & \textbf{S4} & \textbf{S5} & \textbf{S6} & \textbf{S7} & \textbf{S8} & \textbf{Avg} \\
\midrule
\multirow{3}{*}{\textbf{Ours (full)}} 
& Mean Rank $\downarrow$ & 1.03 & 1.12 & 11.25 & 4.69 & 1.77 & 3.09 & 7.60 & 11.67 & 5.28 \\
& R@1 $\uparrow$         & 0.99 & 0.98 & 0.75 & 0.79 & 0.87 & 0.83 & 0.72 & 0.62 & 0.82 \\
& RSA $\uparrow$         & 0.60 & 0.61 & 0.47 & 0.51 & 0.60 & 0.52 & 0.50 & 0.43 & 0.53 \\
\bottomrule
\end{tabular}
\end{table}

\subsection{Within-subject encoder evaluation}

We first evaluate whether the subject-specific encoder extracts stable stimulus representations from noisy repeated fMRI responses, using the within-subject retrieval protocol defined above. Encoder ablations and hyperparameter configurations are provided in Appendix~\ref{sec:ablations}.

\textbf{Performance across subjects.}
Table~\ref{tab:within_subject_per_subject} reports full-encoder performance for each subject. In general, our embeddings are stable: with an average rank of $5.28$, within-subject retrieval works well across subjects. S1 and S2 are nearly perfectly matched across repetitions (Mean Rank $\approx 1$; R@1 $>0.98$), whereas S3 and S8 show lower retrieval performance. Even for higher-performing subjects, RSA remains around $0.6$, indicating that robust instance-level identification does not require exact preservation of the full pairwise geometry.

\textbf{Comparison with encoder baselines.}
Table~\ref{tab:within_subject_main} compares our encoder against three baseline families. First, we include direct neural baselines using preprocessed GLMsingle~\cite{GLMSingle2022} responses and PCA-reduced fMRI responses. Second, we compare against multiview methods trained self-supervised from repeated presentations. Third, we include model-guided baselines, where fMRI responses are linearly regressed to pretrained model embeddings. Direct neural baselines provide limited retrieval, while CCA-based multiview methods improve instance-level matching but yield low RSA. Model-guided baselines preserve stronger geometry, but are weaker at retrieving repeated presentations of the same image. Our full encoder achieves the lowest Mean Rank and highest R@1, with RSA comparable to the strongest model-guided baselines. Thus, repetition-based self-supervision recovers stronger instance-level image information than externally guided encoders, likely because it directly optimizes invariance across neural measurements of the same stimulus rather than fitting an intermediate model space. See Appendix~\ref{sec:details} for implementation details and per-subject results.

\begin{table}[h]
\centering
\caption{\textbf{Within-subject encoder comparison.}
Retrieval and RSA across repeated presentations, averaged across subjects and repetition pairs. Mean Rank and R@1 measure instance-level stability; RSA measures geometry preservation across repetitions. Lower is better for Mean Rank (chance $=258$); higher is better for R@1 (chance $=0.002$) and RSA.}
\label{tab:within_subject_main}
\resizebox{\linewidth}{!}{%
\begin{tabular}{ll|ccc}
\toprule
\textbf{Method} & \textbf{Setting} & \textbf{Mean Rank} $\downarrow$ & \textbf{R@1} $\uparrow$ & \textbf{RSA} $\uparrow$ \\
\midrule
Preprocessed fMRI (GLMsingle \cite{GLMSingle2022}) & GLM (repetitions) & 144.15 & 0.07 & 0.18 \\
\midrule
PCA & Unsupervised & 78.10 & 0.19 & 0.27 \\
PCA + SplitAE \cite{SplitAE} & SSL Repetitions & 93.78 & 0.07 & 0.29 \\
PCA + DeepCCA \cite{DeepCCA} & SSL Repetitions & 23.21 & 0.52 & 0.19 \\
PCA + Multiview CCA \cite{RegMCCA2011} & SSL Repetitions & 19.23 & 0.60 & 0.21 \\
\midrule
DINOv2 ViT-S/14 \cite{dinov2} + Ridge & Vision Model Guided & 24.82	 & 0.30	 & 0.46  \\ 
CLIP ViT-B/16 \cite{Radford2021} + Ridge & Vision Model Guided & 23.23 & 0.26 & 0.50  \\ 
ViT AugReg-L/16 \cite{steiner2022how} + Ridge & Vision Model Guided & 20.30 & 0.32 & 0.53 \\ 
all-MiniLM-L6-v2 \cite{SentenceBert} + Ridge & Language Model Guided & 30.47 & 0.18 & \textbf{0.56} \\ 
\midrule
Ours (full) & SSL Repetitions & \textbf{5.28} & \textbf{0.82} & 0.53 \\
\bottomrule
\end{tabular}
}
\end{table}

\subsection{Pairwise brain-to-brain translation}

We evaluate brain-to-brain translation through cross-subject
image retrieval and representational similarity.
Our method learns orthogonal mappings between independently
trained subject embeddings using unpaired training images
(Subsection~\ref{subsec:pairwise}).
To address initialization sensitivity, we generate 10 candidate
rotations per directed subject pair and jointly select one
per pair by maximizing cycle consistency across subjects,
using the unsupervised criterion detailed in
Appendix~\ref{sec:unsupervised-selection}.

Fig.~\ref{fig:pairwise_alignment} reports retrieval and RSA
for each directed subject pair. The selected mappings recover
accurate cross-subject correspondences, with nearly perfect
retrieval between S1 and S2. Performance varies across pairs,
with lower retrieval between these subjects and S5.

\begin{figure}[t]
    \centering
    \includegraphics[width=\linewidth]{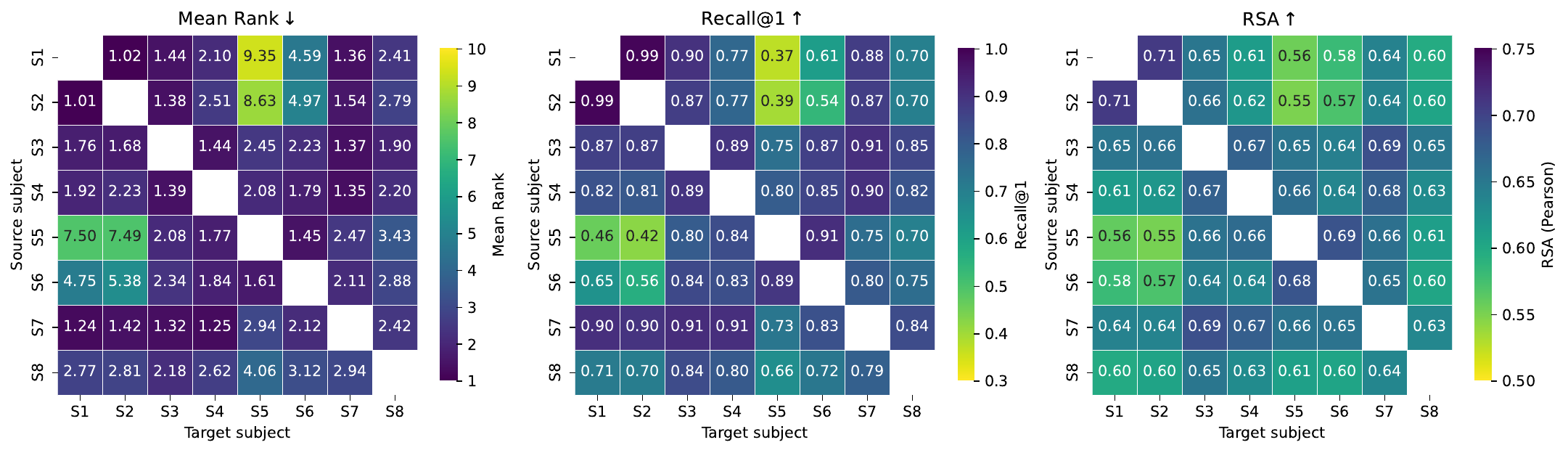}
    \caption{\textbf{Pairwise brain-to-brain translation.}
Performance for each ordered subject pair $(s,t)$. Embeddings from source subject $s$ are mapped into target subject $t$'s space using the unsupervised orthogonal transformation $R_{s \rightarrow t}$ and evaluated on the 515 held-out shared images. Mean Rank (average $\pm$ std: $2.70 \pm 1.84$, chance $=258$) and R@1 (average $\pm$ std: $0.78 \pm 0.14$, chance $=0.002$) measure image-level retrieval after translation. RSA is the Pearson correlation between subject-specific RDMs (average $\pm$ std: $0.63 \pm 0.04$). Darker colors indicate better performance.}
    \label{fig:pairwise_alignment}
\end{figure}

\begin{table}[h]
\centering
\caption{\textbf{Pairwise brain-to-brain translation baselines.}
Performance is reported as the average across ordered off-diagonal subject pairs on the 515 held-out shared images. Rows compare no-translation controls, optimal-transport matching, external model-space references, and our unpaired orthogonal translation. Lower is better for Mean Rank (chance $=258$); higher is better for R@1 (chance $=0.002$) and RSA.}
\label{tab:pairwise_alignment}
\resizebox{\linewidth}{!}{%
    \begin{tabular}{@{}llccc@{}}
\toprule
\textbf{Representation} & \textbf{Alignment}
& \textbf{Mean Rank} $\downarrow$
& \textbf{R@1} $\uparrow$
& \textbf{RSA} $\uparrow$ \\
\midrule

\multicolumn{5}{l}{\textit{No cross-subject alignment}} \\
\addlinespace[2pt]
fMRI betas + PCA
& No alignment & 189.28 & 0.007 & 0.39 \\

\midrule
\multicolumn{5}{l}{\textit{Optimal transport matching on evaluation embeddings (transductive)}} \\
\addlinespace[2pt]
fMRI betas + PCA
& Entropic GW & 166.04 & 0.011 & 0.39 \\
Our embeddings
& Entropic GW & 80.96 & 0.465 & 0.64 \\

\midrule
\multicolumn{5}{l}{\textit{External model-defined shared spaces (unpaired training images)}} \\
\addlinespace[2pt]
ViT AugReg-L/16~\cite{steiner2022how} + Ridge
& Vision model guided & 5.54 & 0.480 & 0.73 \\
CLIP ViT-B/16~\cite{Radford2021} + Ridge
& Vision model guided & 6.60 & 0.440 & 0.70 \\
DINOv2 ViT-S/14~\cite{dinov2} + Ridge
& Vision model guided & 7.00 & 0.470 & 0.67 \\
all-MiniLM-L6-v2~\cite{SentenceBert} + Ridge
& Language model guided & 7.93 & 0.350 & 0.76 \\

\midrule
\multicolumn{5}{l}{\textit{Supervised alignment with paired training images}} \\
\addlinespace[2pt]
Our embeddings
& Ridge regression (CV) & 1.64 & 0.886 & 0.67 \\
Our embeddings
& Orthogonal Procrustes & 1.68 & 0.882 & 0.63 \\
Our embeddings
& CCA & 1.70 & 0.882 & 0.62 \\
Our embeddings
& Linear regression & 3.15 & 0.772 & 0.58 \\

\midrule
\midrule
\multicolumn{5}{l}{\textit{Unsupervised alignment with unpaired training images}} \\
\addlinespace[2pt]
Our embeddings
& Orthogonal Procrustes & 2.70 & 0.780 & 0.63 \\

\bottomrule
\end{tabular}
}
\end{table}

\textbf{Baseline comparisons.}
We compare methods with different sources of alignment information
(Table~\ref{tab:pairwise_alignment}).
The no-alignment control compares PCA-reduced fMRI responses
directly across subjects.
Entropic Gromov--Wasserstein (GW) matching estimates
correspondences from within-subject relational structure,
using the complete evaluation embedding sets without their
identity correspondences.
Model-guided baselines use ridge regression to map each
subject's training responses into a pretrained vision or
language model's feature space. They require no shared
training images across subjects but rely on an external
model to define a common coordinate system.
Supervised Procrustes, cross-validated ridge regression,
CCA, and linear regression quantify the benefit of paired
training images. They use the same neural embeddings and
an additional set of shared images, up to 485 per subject
pair, while retaining the 515-image evaluation set.
The encoders remain unchanged. Appendix~\ref{sec:cross_subject_baselines} provides
implementation details for all baselines.

Our unsupervised mappings outperform the no-alignment,
GW, and model-guided baselines on retrieval, although
model-guided spaces achieve higher RSA (Table \ref{tab:pairwise_alignment}).
Supervised Procrustes, ridge, and CCA further improve retrieval,
with ridge achieving the strongest pairwise results.
This advantage requires paired cross-subject training images
that our unsupervised alignment does not use.

\subsection{A Platonic brain-to-brain translation layer}
\label{sec:shared-alignment-results}
Pairwise translations map individual subject pairs, but do not
guarantee a coherent coordinate system across subjects.
We therefore synchronize the pairwise rotations to recover one
orthogonal transformation $R_s$ per subject
(Subsection~\ref{subsec:global}), mapping all embeddings into
a common latent space (Fig.~\ref{fig:overview}C).

\begin{figure}[t]
    \centering
    \includegraphics[width=0.77\linewidth]{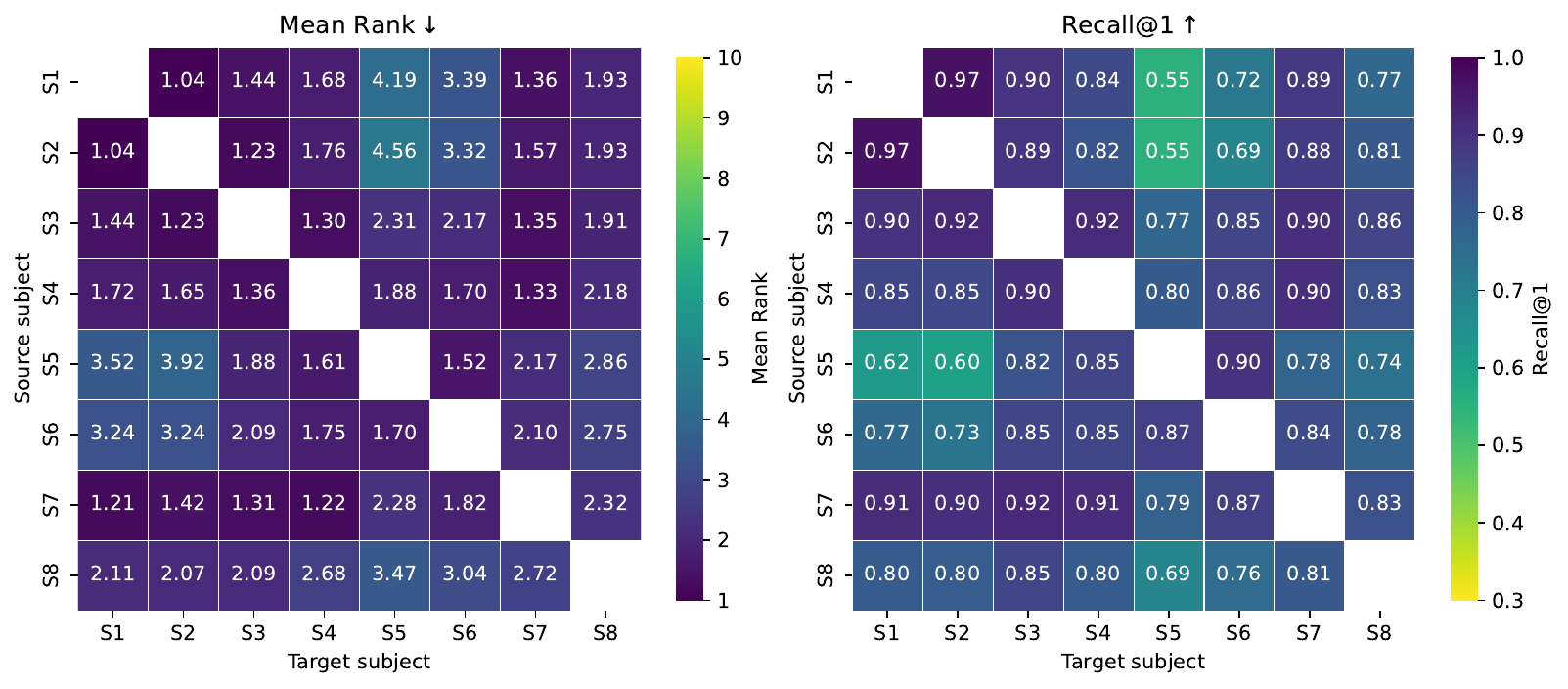}
    \caption{\textbf{Shared-space brain-to-brain translation.}
    Retrieval performance after synchronizing pairwise brain-to-brain rotations into a single shared latent space. Each subject is mapped into the common coordinate system using one orthogonal transformation $R_s$, and retrieval is evaluated across ordered subject pairs on the 515 held-out shared images. Left: Mean Rank (average $\pm$ std: $2.11 \pm 0.84$; chance $=258$). Right: R@1, higher is better (average $\pm$ std: $0.82 \pm 0.09$; chance $=0.002$). Darker colors indicate better performance.}
    \label{fig:global_alignment}
\end{figure}

Synchronization improves average Mean Rank from $2.70$ to
$2.11$ and R@1 from $0.78$ to $0.82$ across the 56 directed
subject pairs (Fig.~\ref{fig:global_alignment}).
Synchronized retrieval exceeded chance in image-identity
permutation tests ($p<10^{-4}$), and remained strong after
excluding each subject's evaluation pairs.
Appendix~\ref{sec:statistical_analysis} reports these analyses
and image-bootstrap confidence intervals.
Appendix~\ref{sec:qualitative_examples} illustrates successful
retrievals, near misses, and failures across subject pairs.

The improvement after synchronization suggests that enforcing
consistency across the subject graph reduces noise in pairwise
estimates. One orthogonal transformation per subject is
sufficient to recover accurate cross-subject correspondence
in a common coordinate system, providing evidence consistent
with the Strong Platonic Representation Hypothesis within
the evaluated visual domain.

\subsection{Model--brain alignment}

We next test whether the final-layer representations of four artificial neural networks predict our fMRI embeddings. For each subject, we fit supervised mappings from model to neural embeddings using the same images used to train the subject encoder. We compare semi-orthogonal mappings, which accommodate differences in embedding dimensionality, with ridge regression. We then map model embeddings of the 515 held-out shared images into the subject's neural embedding space and evaluate retrieval and RSA against the corresponding brain embeddings.

Table~\ref{tab:model_brain_alignment} shows that semi-orthogonal mappings recover moderate instance-level correspondence (best Mean Rank $=13.33$, R@1 $=0.29$).
Ridge regression improves retrieval and RSA (best Mean Rank $=5.86$, R@1 $=0.51$), indicating that model features predict neural embeddings more accurately when more flexible linear transformations are allowed. However, these supervised model-to-brain mappings remain below the shared-space brain-to-brain retrieval results
(Mean Rank $=2.11$, R@1 $=0.82$).

\begin{table}[t]
\centering
\caption{\textbf{Supervised model-to-brain alignment.}
Final-layer model embeddings are mapped to neural embeddings
using each subject's encoder training images and evaluated
on the 515 held-out shared images. We compare semi-orthogonal
mappings with ridge regression. Values are averaged across
subjects. Lower is better for Mean Rank; higher is better
for R@1 and RSA.}
\label{tab:model_brain_alignment}
\resizebox{0.85\linewidth}{!}{%
    \begin{tabular}{lccc|ccc}
\toprule
\multirow{2}{*}{\textbf{Source model}} 
& \multicolumn{3}{c|}{\textbf{Semi-orthogonal mapping}} 
& \multicolumn{3}{c}{\textbf{Ridge regression}} \\
\cmidrule(lr){2-4} \cmidrule(lr){5-7}
& \textbf{Mean Rank} $\downarrow$ & \textbf{R@1} $\uparrow$ & \textbf{RSA} $\uparrow$
& \textbf{Mean Rank} $\downarrow$ & \textbf{R@1} $\uparrow$ & \textbf{RSA} $\uparrow$ \\
\midrule
all-MiniLM-L6-v2 & 35.62 & 0.08 & 0.31 & 27.05 & 0.11 & 0.45 \\
CLIP ViT-B/16 & 21.65 & 0.23 & 0.23 & 7.86 & 0.44 & 0.58 \\
ViT AugReg-L/16 & \textbf{13.33} & \textbf{0.29} & \textbf{0.41} & 8.85 & 0.41 & 0.57 \\
DINOv2 ViT-S/14 & 22.71 & 0.24 & 0.24 & \textbf{5.86} & \textbf{0.51} & \textbf{0.57} \\
\bottomrule
\end{tabular}
}
\end{table}

\subsection{Cross-subject fMRI reconstruction}
\label{sec:cross-subject-reconstruction}

Finally, we test whether aligned embeddings preserve information
for predicting another subject's voxel responses.
For each target subject $t$, we fit a ridge decoder
$D_t(z)=zW_t+b_t$ from its shared-space training embeddings
$Z_{\mathrm{train}}^{(t)}R_t$ to its measured voxel responses.
We then apply this decoder, without refitting, to another
subject's aligned test embeddings:
\begin{equation}
\widehat{X}_{s\rightarrow t}
=
D_t\!\left(Z_{\mathrm{test}}^{(s)}R_s\right).
\end{equation}
The rotations are learned unsupervised, and the 515 shared
evaluation images are excluded from decoder training.

We train target decoders for S1, S2, S5, and S7, who completed
all NSD sessions, and use all eight subjects as embedding
sources. Embeddings and voxel responses are averaged across
repetitions. We evaluate reconstruction using Pearson
correlation across the evaluation images for each voxel,
then average correlations across target voxels.

Across the 28 cross-subject evaluations, mean correlation (average $\pm$ SD)
was $0.350 \pm 0.038$, compared with $0.431 \pm 0.046$
for within-subject reconstruction
(Table~\ref{tab:cross_subject_reconstruction}).
Cross-subject prediction therefore retained approximately
81\% of the within-subject correlation, showing that the
shared embeddings contain transferable information about
measured neural responses.

\begin{table}[ht]
\centering
\small
\caption{\textbf{Cross-subject fMRI reconstruction.}
Mean voxel-wise Pearson correlations between predicted and
measured responses across 515 held-out images.
Rows denote target subjects and columns denote embedding
sources. Bold diagonal entries report within-subject
reconstruction. Each target decoder is trained only on
that subject's training data and applied without refitting.}
\label{tab:cross_subject_reconstruction}
\begin{tabular}{@{}lcccccccc@{}}
\toprule
\textbf{Target / Source}
& \textbf{S1} & \textbf{S2} & \textbf{S3} & \textbf{S4}
& \textbf{S5} & \textbf{S6} & \textbf{S7} & \textbf{S8} \\
\midrule
\textbf{S1}
& \textbf{0.433} & 0.364 & 0.350 & 0.340
& 0.328 & 0.334 & 0.348 & 0.341 \\
\textbf{S2}
& 0.372 & \textbf{0.428} & 0.356 & 0.348
& 0.336 & 0.340 & 0.349 & 0.346 \\
\textbf{S5}
& 0.388 & 0.390 & 0.410 & 0.417
& \textbf{0.489} & 0.416 & 0.406 & 0.399 \\
\textbf{S7}
& 0.309 & 0.309 & 0.306 & 0.305
& 0.298 & 0.301 & \textbf{0.376} & 0.295 \\
\bottomrule
\end{tabular}
\end{table}

\section{Related work}

\textbf{Universal geometry in machine learning.}
Learned representations in deep neural networks exhibit structured geometries that reflect statistical properties of the data. Empirical work shows that models trained with different architectures, objectives, and modalities often develop similar representational spaces~\cite{Li2016,mikolov2013,Raghu2017,Kornblith2019}. These observations motivate the \textit{Platonic Representation Hypothesis}~\cite{Huh2024}, which proposes that models converge toward a shared latent structure as scale and data coverage increase. This convergence has been studied through representational similarity measures~\cite{Kornblith2019} and constructive methods such as model stitching and latent-space mapping~\cite{bansal2021revisiting,AlvarezMelis2018,Grave2019,Moschella2022}. Recent work tests a stronger version of this hypothesis by showing that embedding spaces can be translated from intrinsic geometry alone, without shared inputs or paired supervision~\cite{Rishi2025,lample2018word,Dar2026}. In this work, we ask whether this constructive form---the Strong Platonic Representation Hypothesis---also holds for biological neural representations.

\textbf{Shared representations in neuroscience.}
Shared neural representations are harder to recover because brain data are noisy, limited, and subject to anatomical and functional variability. Nevertheless, inter-subject synchrony~\cite{Hasson2004,Nastase2019} and representational similarity analyses~\cite{Kriegeskorte2008,Lin2024} show that neural response structure is partly preserved across subjects~\cite{MarcosManchon2025,Chen2017,Diedrichsen2017}. Functional alignment methods such as hyperalignment map subjects into common spaces, but typically require shared stimuli to establish correspondences~\cite{Haxby2011,Guntupalli2016}. Other methods introduce anchors through model-derived feature spaces~\cite{zhou2024clipmused} or image-to-fMRI encoders~\cite{Wasserman2026}, while unsupervised alternatives infer correspondences through structural or distributional matching~\cite{Nakamura2024} and optimal transport-based functional alignment~\cite{Bazeille2019,Bazeille2021}. In contrast, we learn subject-specific fMRI embeddings from neural repetitions and test whether they can be translated by orthogonal rotations learned from unpaired subject-specific images.

\textbf{Model--brain alignment.}
A complementary literature aligns neural activity with deep neural network representations, revealing systematic correspondences between cortical processing stages and model layers~\cite{Yamins2014,Khaligh2014,Cichy2016,MarcosManchon2025}. These approaches support encoding, decoding, and benchmarking efforts such as \emph{Brain-Score}~\cite{Schrimpf2020,Caucheteux2022,Konkle2022,Wang2023,Doerig2025,Conwell2024}. Recent work further suggests that brains and models may share low-dimensional representational axes~\cite{Chen2025}. However, model-mediated approaches impose or evaluate neural representations through an external feature geometry. Our method instead recovers a shared brain space directly from neural data, using model--brain mappings only diagnostically.

\section{Discussion}

The Strong Platonic Representation Hypothesis \cite{Rishi2025} posits that the universal latent structure of representations not only exists but can be harnessed to translate representations from one space to another without any paired data or model access. Whereas the original hypothesis was conjectured over artificial embeddings, in this work, we evaluated this prediction in embeddings learned from human fMRI responses to natural images. We found that subject-specific embeddings learned from brain data alone can be translated across individuals using unsupervised orthogonal rotations, without paired cross-subject samples or intermediate model representations. Synchronizing these rotations into a shared latent space further improved retrieval, indicating that the pairwise translations are mutually compatible with a common coordinate system. Together, these results suggest that independently learned subject embeddings share a geometric structure that supports cross-subject translation within the evaluated visual domain.

This shared representational structure has practical implications for cross-subject neural modeling. In our fMRI reconstruction experiment, a target-subject ridge decoder predicts voxel responses from another subject's aligned embeddings without refitting (Table~\ref{tab:cross_subject_reconstruction}). Such transfer could reduce subject-specific calibration for image-to-fMRI encoding, fMRI-to-image decoding, and synthetic neural data generation. Shared spaces could also support training neural models on heterogeneous datasets in which subjects viewed different stimuli.

Our model--brain results further distinguish linear predictability
from geometric alignment. Model features predict neural embeddings
under supervised linear mappings, with ridge regression outperforming
semi-orthogonal mappings. This suggests that current models capture
stimulus information relevant to neural responses, while more
flexible transformations help align their representational spaces.
Brain-derived shared spaces may therefore provide a useful target
for investigating which aspects of representation are shared
or differ between artificial models and neural embeddings.

Several limitations remain. Our geometric conclusions concern embeddings learned with a common encoder architecture and training objective, whose contributions to the observed alignment remain to be disentangled. Establishing approximate isometry would also require direct tests of distance preservation beyond retrieval performance. The analysis relies on high-quality fMRI data with repeated stimulus presentations to isolate stable stimulus-related signal from trial-level noise. It remains unclear whether comparable alignment will hold in lower-SNR modalities, smaller datasets, or higher-level cognitive tasks. Future work should test whether shared representational structures extend to more distant stimulus distributions, task contexts, and clinical datasets. Finally, subject-agnostic neural spaces may improve decoding and transfer learning, but related methods raise neural privacy concerns if applied to sensitive brain data without consent; future use should require explicit consent, careful data governance, and safeguards against misuse.

\begin{ack}
This work was supported by the Spanish Ministerio de Ciencia, Innovación y Universidades through the Agencia Estatal de Investigación (AEI), under project PID2022-140426NB-I00, co-funded by the European Regional Development Fund (ERDF, ``A way to build Europe''). We thank the CERCA Programme/Generalitat de Catalunya for institutional support. R.J. was supported by the Digital Life Initiative Doctoral Fellowship at Cornell Tech.
\end{ack}

{
\small
\bibliographystyle{unsrtnat}
\bibliography{references}

@string{j_NatNeuro = "Nature Neuroscience"}

@string{c_ICML = "International Conference on Machine Learning (ICML)"}

@string{j_PNAS = "Proceedings of the National Academy of Sciences"}

@string{j_FrontSystNeuro = "Frontiers in Systems Neuroscience"}

@string{j_Science = "Science"}

@string{j_SciRep = "Scientific Reports"}

@string{j_NatCommun = "Nature Communications"}

@string{j_eLife = "eLife"}

@string{j_PLoSCompBiol = "PLOS Computational Biology"}

@string{j_Neuron = "Neuron"}

@string{j_CommBiol = "Communications Biology"}

@string{j_NatMachineInt = "Nature Machine Intelligence"}

@string{j_TMLR = "Transactions on Machine Learning Research"}

@string{j_Nature = "Nature"}

@string{c_CVPR = "IEEE/CVF Conference on Computer Vision and Pattern Recognition (CVPR)"}

@string{c_NeurIPS = "Advances in Neural Information Processing Systems"}

@Article{MarcosManchon2025,
    author={Marcos-Manch{\'o}n, Pablo
    and Fuentemilla, Llu{\'i}s},
    title={Shared representations in brains and models reveal a two-route cortical organization during scene perception},
    journal={Communications Biology},
    year={2026},
    month={May},
    doi={10.1038/s42003-026-10169-0},
}

@inproceedings{Rishi2025,
    title={Harnessing the Universal Geometry of Embeddings},
    author={Rishi Dev Jha and Collin Zhang and Vitaly Shmatikov and John Xavier Morris},
    booktitle = c_NeurIPS,
    year={2025},
}

@misc{Dar2026,
      title={{mini-vec2vec}: Scaling Universal Geometry Alignment with Linear Transformations}, 
      author={Guy Dar},
      year={2026},
      eprint={2510.02348},
      archivePrefix={arXiv},
      primaryClass={cs.CL},
      url={https://arxiv.org/abs/2510.02348}, 
}

@inproceedings{Huh2024,
author = {Huh, Minyoung and Cheung, Brian and Wang, Tongzhou and Isola, Phillip},
title = {{The platonic representation hypothesis}},
year = {2024},
publisher = {JMLR.org},
booktitle = {Proceedings of the 41st International Conference on Machine Learning},
articleno = {827},
numpages = {26},
}

@article{Liu2016,
title = {Noise contributions to the {fMRI} signal: An overview},
journal = {NeuroImage},
volume = {143},
pages = {141-151},
year = {2016},
doi = {10.1016/j.neuroimage.2016.09.008},
author = {Thomas T. Liu},
}

@Article{Ozcelik2023,
author={Ozcelik, Furkan
and VanRullen, Rufin},
title={Natural scene reconstruction from {fMRI} signals using generative latent diffusion},
journal={Scientific Reports},
year={2023},
month={Sep},
day={20},
volume={13},
number={1},
pages={15666},
doi={10.1038/s41598-023-42891-8},
}

@inproceedings{
scotti2023reconstructing,
title={Reconstructing the Mind's Eye: f{MRI}-to-Image with Contrastive Learning and Diffusion Priors},
author={Paul Steven Scotti and Atmadeep Banerjee and Jimmie Goode and Stepan Shabalin and Alex Nguyen and Cohen Ethan and Aidan James Dempster and Nathalie Verlinde and Elad Yundler and David Weisberg and Kenneth Norman and Tanishq Mathew Abraham},
booktitle={Thirty-seventh Conference on Neural Information Processing Systems},
year={2023},
url={https://openreview.net/forum?id=rwrblCYb2A}
}

@article{MCCA1971,
    author = {Kettenring, J. R.},
    title = {Canonical analysis of several sets of variables},
    journal = {Biometrika},
    volume = {58},
    number = {3},
    pages = {433-451},
    year = {1971},
    doi = {10.1093/biomet/58.3.433},
}

@article{RegMCCA2011,
  title   = {Regularized generalized canonical correlation analysis},
  author  = {Tenenhaus, Arthur and Tenenhaus, Michel},
  journal = {Psychometrika},
  year    = {2011},
  volume  = {76},
  number  = {2},
  pages   = {257--284}
}

@misc{infoNCE2019,
      title={Representation Learning with Contrastive Predictive Coding}, 
      author={Aaron van den Oord and Yazhe Li and Oriol Vinyals},
      year={2019},
      eprint={1807.03748},
      archivePrefix={arXiv},
      primaryClass={cs.LG},
      url={https://arxiv.org/abs/1807.03748}, 
}

@article{Kriegeskorte2008,
  author  = {Kriegeskorte, Nikolaus and Mur, Marieke and Bandettini, Peter A.},
  title   = {Representational similarity analysis -- connecting the branches of systems neuroscience},
  journal = j_FrontSystNeuro,
  volume  = {2},
  year    = {2008},
  doi     = {10.3389/neuro.06.004.2008}
}

@article{ICP,
author = {Besl, Paul J. and McKay, Neil D.},
title = {A Method for Registration of {3-D} Shapes},
year = {1992},
publisher = {IEEE Computer Society},
volume = {14},
number = {2},
doi = {10.1109/34.121791},
journal = {IEEE Trans. Pattern Anal. Mach. Intell.},
}

@article{Singer2011,
title = {Angular synchronization by eigenvectors and semidefinite programming},
journal = {Applied and Computational Harmonic Analysis},
volume = {30},
number = {1},
pages = {20-36},
year = {2011},
doi = {10.1016/j.acha.2010.02.001},
author = {A. Singer},
}

@article{WangRotation2013,
author = {Wang, Lanhui and Singer, Amit},
year = {2013},
month = {10},
pages = {145-193},
title = {Exact and Stable Recovery of Rotations for Robust Synchronization},
volume = {2},
journal = {Information and Inference: A Journal of the IMA},
doi = {10.1093/imaiai/iat005}
}

@article{Allen2022,
  author  = {Allen, Emily J. and St-Yves, Ghislain and Wu, Yihan and Breedlove, Jesse L. and Prince, Jacob S. and Dowdle, Logan T. and Nau, Matthias and Caron, Brad and Pestilli, Franco and Charest, Ian and Hutchinson, J. Benjamin and Naselaris, Thomas and Kay, Kendrick},
  title   = {A massive {7T} {fMRI} dataset to bridge cognitive neuroscience and artificial intelligence},
  journal = j_NatNeuro,
  volume  = {25},
  number  = {1},
  pages   = {116--126},
  year    = {2022},
  doi     = {10.1038/s41593-021-00962-x}
}

@inproceedings{Lin2014MicrosoftCC,
  title={{Microsoft COCO}: Common Objects in Context},
  author={Tsung-Yi Lin and Michael Maire and Serge J. Belongie and James Hays and Pietro Perona and Deva Ramanan and Piotr Doll{\'a}r and C. Lawrence Zitnick},
  booktitle={European Conference on Computer Vision (ECCV)},
  year={2014},
  pages={740--755}
}

@inproceedings{Raghu2017,
author = {Raghu, Maithra and Gilmer, Justin and Yosinski, Jason and Sohl-Dickstein, Jascha},
title = {{SVCCA}: singular vector canonical correlation analysis for deep learning dynamics and interpretability},
year = {2017},
isbn = {9781510860964},
publisher = {Curran Associates Inc.},
address = {Red Hook, NY, USA},
booktitle = {Proceedings of the 31st International Conference on Neural Information Processing Systems},
pages = {6078–6087},
numpages = {10},
location = {Long Beach, California, USA},
series = {NIPS'17}
}

@inproceedings{Kornblith2019,
  author    = {Kornblith, Simon and Norouzi, Mohammad and Lee, Honglak and Hinton, Geoffrey},
  title     = {Similarity of neural network representations revisited},
  booktitle = c_ICML,
  year      = {2019},
  volume    = {97},
  pages     = {3519--3529}
}

@article{
sucholutsky2025,
title={Getting aligned on representational alignment},
author={Ilia Sucholutsky and Lukas Muttenthaler and Adrian Weller and Andi Peng and Andreea Bobu and Been Kim and Bradley C. Love and Christopher J Cueva and Erin Grant and Iris Groen and Jascha Achterberg and Joshua B. Tenenbaum and Katherine M. Collins and Katherine Hermann and Kerem Oktar and Klaus Greff and Martin N Hebart and Nathan Cloos and Nikolaus Kriegeskorte and Nori Jacoby and Qiuyi Zhang and Raja Marjieh and Robert Geirhos and Sherol Chen and Simon Kornblith and Sunayana Rane and Talia Konkle and Thomas O'Connell and Thomas Unterthiner and Andrew Kyle Lampinen and Klaus Robert Muller and Mariya Toneva and Thomas L. Griffiths},
journal={Transactions on Machine Learning Research},
issn={2835-8856},
year={2025},
url={https://openreview.net/forum?id=Hiq7lUh4Yn},
}

@inproceedings{
lample2018word,
title={Word translation without parallel data},
author={Guillaume Lample and Alexis Conneau and Marc'Aurelio Ranzato and Ludovic Denoyer and Hervé Jégou},
booktitle={International Conference on Learning Representations},
year={2018},
url={https://openreview.net/forum?id=H196sainb},
}

@inproceedings{lu2025survey,
    title = "Representation Potentials of Foundation Models for Multimodal Alignment: A Survey",
    author = "Lu, Jianglin  and
      Wang, Hailing  and
      Xu, Yi  and
      Wang, Yizhou  and
      Yang, Kuo  and
      Fu, Yun",
    booktitle = "Proceedings of the 2025 Conference on Empirical Methods in Natural Language Processing",
    year = "2025",
    publisher = "Association for Computational Linguistics",
    doi = "10.18653/v1/2025.emnlp-main.843",
    pages = "16669--16684",
    ISBN = "979-8-89176-332-6",
}

@inproceedings{
bansal2021revisiting,
title={Revisiting Model Stitching to Compare Neural Representations},
author={Yamini Bansal and Preetum Nakkiran and Boaz Barak},
booktitle={Advances in Neural Information Processing Systems},
editor={A. Beygelzimer and Y. Dauphin and P. Liang and J. Wortman Vaughan},
year={2021},
url={https://openreview.net/forum?id=ak06J5jNR4}
}

@inproceedings{mikolov2013,
    title = "Linguistic Regularities in Continuous Space Word Representations",
    author = "Mikolov, Tomas  and
      Yih, Wen-tau  and
      Zweig, Geoffrey",
    booktitle = "Proceedings of the 2013 Conference of the North {A}merican Chapter of the Association for Computational Linguistics: Human Language Technologies",
    year = "2013",
    publisher = "Association for Computational Linguistics",
    pages = "746--751"
}

@article{Hasson2004,
  author  = {Hasson, Uri and Nir, Yuval and Levy, Ifat and Fuhrmann, Galit and Malach, Rafael},
  title   = {Intersubject synchronization of cortical activity during natural vision},
  journal = j_Science,
  volume  = {303},
  number  = {5664},
  pages   = {1634--1640},
  year    = {2004},
  doi     = {10.1126/science.1089506}
}

@article{Nastase2019,
author={Nastase, Samuel A
and Gazzola, Valeria
and Hasson, Uri
and Keysers, Christian},
title={Measuring shared responses across subjects using intersubject correlation},
journal={Social Cognitive and Affective Neuroscience},
year={2019},
volume={14},
number={6},
pages={667-685},
doi={10.1093/scan/nsz037},
}

@article{Haxby2011,
title = {A Common, High-Dimensional Model of the Representational Space in Human Ventral Temporal Cortex},
journal = {Neuron},
volume = {72},
number = {2},
pages = {404-416},
year = {2011},
doi = {10.1016/j.neuron.2011.08.026},
author = {James V. Haxby and J. Swaroop Guntupalli and Andrew C. Connolly and Yaroslav O. Halchenko and Bryan R. Conroy and M. Ida Gobbini and Michael Hanke and Peter J. Ramadge},
}

@article{Guntupalli2016,
    author = {Guntupalli, J. Swaroop and Hanke, Michael and Halchenko, Yaroslav O. and Connolly, Andrew C. and Ramadge, Peter J. and Haxby, James V.},
    title = {A Model of Representational Spaces in Human Cortex},
    journal = {Cerebral Cortex},
    volume = {26},
    number = {6},
    pages = {2919-2934},
    year = {2016},
    doi = {10.1093/cercor/bhw068},
}

@article{Yamins2014,
  author  = {Yamins, Daniel L. K. and Hong, Ha and Cadieu, Charles F. and Solomon, Ethan A. and Seibert, Darren and DiCarlo, James J.},
  title   = {Performance-optimized hierarchical models predict neural responses in higher visual cortex},
  journal = j_PNAS,
  volume  = {111},
  number  = {23},
  pages   = {8619--8624},
  year    = {2014},
  doi     = {10.1073/pnas.1403112111}
}

@article{Cichy2016,
  author  = {Cichy, Radoslaw Martin and Khosla, Aditya and Pantazis, Dimitrios and Torralba, Antonio and Oliva, Aude},
  title   = {Comparison of deep neural networks to spatio-temporal cortical dynamics of human visual object recognition reveals hierarchical correspondence},
  journal = j_SciRep,
  volume  = {6},
  number  = {1},
  pages   = {27755},
  year    = {2016},
  doi     = {10.1038/srep27755}
}

@article{Chen2025,
  author  = {Chen, Zirui and Bonner, Michael F.},
  title   = {Universal dimensions of visual representation},
  journal = {Science Advances},
  volume  = {11},
  number  = {27},
  pages   = {eadw7697},
  year    = {2025},
  doi     = {10.1126/sciadv.adw7697}
}

@article{Conwell2024,
  author  = {Conwell, Colin and Prince, Jacob S. and Kay, Kendrick N. and Alvarez, George A. and Konkle, Talia},
  title   = {A large-scale examination of inductive biases shaping high-level visual representation in brains and machines},
  journal = j_NatCommun,
  volume  = {15},
  number  = {1},
  pages   = {9383},
  year    = {2024},
  doi     = {10.1038/s41467-024-53147-y}
}

@article{Lin2024,
author = {Baihan Lin  and Nikolaus Kriegeskorte },
title = {The topology and geometry of neural representations},
journal = {Proceedings of the National Academy of Sciences},
volume = {121},
number = {42},
pages = {e2317881121},
year = {2024},
doi = {10.1073/pnas.2317881121},
URL = {https://www.pnas.org/doi/abs/10.1073/pnas.2317881121},
}

@article{GLMSingle2022,
  author  = {Prince, Jacob S. and Charest, Ian and Kurzawski, Jan W. and Pyles, John A. and Tarr, Michael J. and Kay, Kendrick N.},
  title   = {Improving the accuracy of single-trial {fMRI} response estimates using {GLMsingle}},
  journal = j_eLife,
  volume  = {11},
  year    = {2022},
  doi     = {10.7554/eLife.77599}
}

@InProceedings{Li2016,
  title = 	 {Convergent Learning: Do different neural networks learn the same representations?},
  author = 	 {Li, Yixuan and Yosinski, Jason and Clune, Jeff and Lipson, Hod and Hopcroft, John},
  booktitle = 	 {Proceedings of the 1st International Workshop on Feature Extraction: Modern Questions and Challenges at NIPS 2015},
  pages = 	 {196--212},
  year = 	 {2015},
  volume = 	 {44},
  series = 	 {Proceedings of Machine Learning Research},
  publisher =    {PMLR},
}

@inproceedings{AlvarezMelis2018,
    title = "{G}romov-{W}asserstein Alignment of Word Embedding Spaces",
    author = "Alvarez-Melis, David  and
      Jaakkola, Tommi",
    booktitle = "Proceedings of the 2018 Conference on Empirical Methods in Natural Language Processing",
    year = "2018",
    publisher = "Association for Computational Linguistics",
    doi = "10.18653/v1/D18-1214",
    pages = "1881--1890",
}

@InProceedings{Grave2019,
  title = 	 {Unsupervised Alignment of Embeddings with {Wasserstein} {Procrustes}},
  author =       {Grave, Edouard and Joulin, Armand and Berthet, Quentin},
  booktitle = 	 {Proceedings of the Twenty-Second International Conference on Artificial Intelligence and Statistics},
  pages = 	 {1880--1890},
  year = 	 {2019},
  volume = 	 {89},
  series = 	 {Proceedings of Machine Learning Research},
}

@inproceedings{Moschella2022,
title={Relative representations enable zero-shot latent space communication},
author={Luca Moschella and Valentino Maiorca and Marco Fumero and Antonio Norelli and Francesco Locatello and Emanuele Rodol{\`a}},
booktitle={The Eleventh International Conference on Learning Representations },
year={2023},
url={https://openreview.net/forum?id=SrC-nwieGJ}
}

@article{Chen2017,
  author  = {Chen, Janice and Leong, Yuan Chang and Honey, Christopher J. and Yong, Chung H. and Norman, Kenneth A. and Hasson, Uri},
  title   = {Shared memories reveal shared structure in neural activity across individuals},
  journal = j_NatNeuro,
  volume  = {20},
  number  = {1},
  pages   = {115--125},
  year    = {2017},
  doi     = {10.1038/nn.4450}
}

@article{Diedrichsen2017,
author={Diedrichsen, Jörn and Kriegeskorte, Nikolaus},
title={Representational models: A common framework for understanding encoding, pattern-component, and representational-similarity analysis},
journal={PLOS Computational Biology},
year={2017},
volume={13},
number={4},
pages={1-33},
doi={10.1371/journal.pcbi.1005508},
}

@article{Khaligh2014,
  author  = {Khaligh-Razavi, Seyed-Mahdi and Kriegeskorte, Nikolaus},
  title   = {Deep supervised, but not unsupervised, models may explain {IT} cortical representation},
  journal = j_PLoSCompBiol,
  volume  = {10},
  number  = {11},
  pages   = {1--29},
  year    = {2014},
  doi     = {10.1371/journal.pcbi.1003915}
}

@article{Schrimpf2020,
  author  = {Schrimpf, Martin and Kubilius, Jonas and Lee, Michael J. and Ratan Murty, N. Apurva and Ajemian, Robert and DiCarlo, James J.},
  title   = {Integrative benchmarking to advance neurally mechanistic models of human intelligence},
  journal = j_Neuron,
  volume  = {108},
  number  = {3},
  pages   = {413--423},
  year    = {2020},
  doi     = {10.1016/j.neuron.2020.07.040}
}

@article{Caucheteux2022,
  author  = {Caucheteux, Charlotte and King, Jean-R{\'e}mi},
  title   = {Brains and algorithms partially converge in natural language processing},
  journal = j_CommBiol,
  volume  = {5},
  number  = {1},
  pages   = {134},
  year    = {2022},
  doi     = {10.1038/s42003-022-03036-1}
}

@article{Konkle2022,
author={Konkle, Talia
and Alvarez, George A.},
title={A self-supervised domain-general learning framework for human ventral stream representation},
journal={Nature Communications},
year={2022},
volume={13},
number={1},
pages={491},
doi={10.1038/s41467-022-28091-4},
}

@article{Wang2023,
author={Wang, Aria Y.
and Kay, Kendrick
and Naselaris, Thomas
and Tarr, Michael J.
and Wehbe, Leila},
title={Better models of human high-level visual cortex emerge from natural language supervision with a large and diverse dataset},
journal=j_NatMachineInt,
year={2023},
volume={5},
number={12},
pages={1415-1426},
doi={10.1038/s42256-023-00753-y},
}

@article{Doerig2025,
author={Doerig, Adrien
and Kietzmann, Tim C.
and Allen, Emily
and Wu, Yihan
and Naselaris, Thomas
and Kay, Kendrick
and Charest, Ian},
title={High-level visual representations in the human brain are aligned with large language models},
journal=j_NatMachineInt,
year={2025},
pages={1220-1234},
doi={10.1038/s42256-025-01072-0},
}

@inproceedings{
    zhou2024clipmused,
    title={{CLIP}-{MUSED}: {CLIP}-Guided Multi-Subject Visual Neural Information Semantic Decoding},
    author={Qiongyi Zhou and Changde Du and Shengpei Wang and Huiguang He},
    booktitle={The Twelfth International Conference on Learning Representations},
    year={2024},
    url={https://openreview.net/forum?id=lKxL5zkssv}
}

@inproceedings{Radford2021,
  author    = {Radford, Alec and Kim, Jong Wook and Hallacy, Chris and Ramesh, Aditya and Goh, Gabriel and Agarwal, Sandhini and Sastry, Girish and Askell, Amanda and Mishkin, Pamela and Clark, Jack and Krueger, Gretchen and Sutskever, Ilya},
  title     = {Learning transferable visual models from natural language supervision},
  booktitle = c_ICML,
  year      = {2021},
  pages     = {8748--8763},
}

@inproceedings{DeepCCA,
  title = 	 {Deep Canonical Correlation Analysis},
  author = 	 {Andrew, Galen and Arora, Raman and Bilmes, Jeff and Livescu, Karen},
  booktitle = {Proceedings of the 30th International Conference on Machine Learning},
  pages = 	 {1247--1255},
  year = 	 {2013},
  volume = 	 {28},
  url = 	 {https://proceedings.mlr.press/v28/andrew13.html},
}

@article{dinov2,
  author  = {Oquab, Maxime and Darcet, Timoth{\'e}e and Moutakanni, Th{\'e}o and Vo, Huy V. and Szafraniec, Marc and Khalidov, Vasil and Fernandez, Pierre and Haziza, Daniel and Massa, Francisco and El-Nouby, Alaaeldin and Assran, Mido and Ballas, Nicolas and Galuba, Wojciech and Howes, Russell and Huang, Po-Yao and Li, Shang-Wen and Misra, Ishan and Rabbat, Michael and Sharma, Vasu and Synnaeve, Gabriel and Xu, Hu and Jegou, Herve and Mairal, Julien and Labatut, Patrick and Joulin, Armand and Bojanowski, Piotr},
  title   = {{DINO}v2: Learning robust visual features without supervision},
  journal = j_TMLR,
  year    = {2024},
  url={https://openreview.net/forum?id=a68SUt6zFt},
}

@article{steiner2022how,
  author  = {Steiner, Andreas Peter and Kolesnikov, Alexander and Zhai, Xiaohua and Wightman, Ross and Uszkoreit, Jakob and Beyer, Lucas},
  title   = {How to train your {ViT}? Data, augmentation, and regularization in vision transformers},
  journal = j_TMLR,
  year    = {2022},
  url={https://openreview.net/forum?id=4nPswr1KcP},
}

@inproceedings{SentenceBert,
    title = "Sentence-{BERT}: Sentence Embeddings using {S}iamese {BERT}-Networks",
    author = "Reimers, Nils  and
      Gurevych, Iryna",
    booktitle = "Proceedings of the 2019 Conference on Empirical Methods in Natural Language Processing and the 9th International Joint Conference on Natural Language Processing (EMNLP-IJCNLP)",
    year = "2019",
    doi = "10.18653/v1/D19-1410",
    pages = "3982--3992",
}

@InProceedings{SplitAE,
  title = 	 {On Deep Multi-View Representation Learning},
  author = 	 {Wang, Weiran and Arora, Raman and Livescu, Karen and Bilmes, Jeff},
  booktitle = 	 {Proceedings of the 32nd International Conference on Machine Learning},
  pages = 	 {1083--1092},
  year = 	 {2015},
  volume = 	 {37},
  url = 	 {https://proceedings.mlr.press/v37/wangb15.html},
}

@ARTICLE{Nakamura2024,
AUTHOR={Nakamura, Daiki  and Kaji, Shizuo  and Kanai, Ryota  and Hayashi, Ryusuke },
TITLE={Unsupervised method for representation transfer from one brain to another},
JOURNAL={Frontiers in Neuroinformatics},
VOLUME={Volume 18 - 2024},
YEAR={2024},
DOI={10.3389/fninf.2024.1470845},
}

@article{Wasserman2026,
title = {Functional brain-to-brain transformation without shared stimuli},
journal = {NeuroImage},
volume = {327},
pages = {121741},
year = {2026},
doi = {10.1016/j.neuroimage.2026.121741},
author = {Navve Wasserman and Roman Beliy and Roy Urbach and Michal Irani},
}

@InProceedings{Bazeille2019,
author="Bazeille, T.
and Richard, H.
and Janati, H.
and Thirion, B.",
editor="Chung, Albert C. S.
and Gee, James C.
and Yushkevich, Paul A.
and Bao, Siqi",
title="Local Optimal Transport for Functional Brain Template Estimation",
booktitle="Information Processing in Medical Imaging",
year="2019",
publisher="Springer International Publishing",
pages="237--248",
}

@article{Bazeille2021,
title = {An empirical evaluation of functional alignment using inter-subject decoding},
journal = {NeuroImage},
volume = {245},
pages = {118683},
year = {2021},
doi = {10.1016/j.neuroimage.2021.118683},
author = {Thomas Bazeille and Elizabeth DuPre and Hugo Richard and Jean-Baptiste Poline and Bertrand Thirion},
}

@article{Kastner2015,
    author = {Wang, Liang and Mruczek, Ryan E.B. and Arcaro, Michael J. and Kastner, Sabine},
    title = {Probabilistic Maps of Visual Topography in Human Cortex},
    journal = {Cerebral Cortex},
    volume = {25},
    number = {10},
    pages = {3911-3931},
    year = {2015},
    doi = {10.1093/cercor/bhu277},
}

@article{Glasser2016,
  author  = {Glasser, Matthew F. and Coalson, Timothy S. and Robinson, Emma C. and Hacker, Carl D. and Harwell, John and Yacoub, Essa and Ugurbil, Kamil and Andersson, Jesper and Beckmann, Christian F. and Jenkinson, Mark and Smith, Stephen M. and Van Essen, David C.},
  title   = {A multi-modal parcellation of human cerebral cortex},
  journal = j_Nature,
  volume  = {536},
  number  = {7615},
  pages   = {171--178},
  year    = {2016},
  doi     = {10.1038/nature18933}
}

@inproceedings{mae,
  author    = {He, Kaiming and Chen, Xinlei and Xie, Saining and Li, Yanghao and Doll{\'a}r, Piotr and Girshick, Ross},
  title     = {Masked autoencoders are scalable vision learners},
  booktitle = c_CVPR,
  year      = {2022},
  pages     = {15979--15988},
  doi       = {10.1109/CVPR52688.2022.01553}
}

@misc{rw2019timm,
  author       = {Wightman, Ross},
  title        = {{PyTorch} image models},
  year         = {2019},
  doi         = {10.5281/zenodo.4414861}
}

@article{Wright2015,
  author  = {Wright, Stephen J.},
  title   = {Coordinate Descent Algorithms},
  journal = {Mathematical Programming},
  year    = {2015},
  volume  = {151},
  number  = {1},
  pages   = {3--34},
  doi     = {10.1007/s10107-015-0892-3},
  url     = {https://doi.org/10.1007/s10107-015-0892-3}
}

@article{flamary2021pot,
  author  = {R{\'e}mi Flamary and Nicolas Courty and Alexandre Gramfort and Mokhtar Z. Alaya and Aur{\'e}lie Boisbunon and Stanislas Chambon and Laetitia Chapel and Adrien Corenflos and Kilian Fatras and Nemo Fournier and L{\'e}o Gautheron and Nathalie T.H. Gayraud and Hicham Janati and Alain Rakotomamonjy and Ievgen Redko and Antoine Rolet and Antony Schutz and Vivien Seguy and Danica J. Sutherland and Romain Tavenard and Alexander Tong and Titouan Vayer},
  title   = {{POT}: Python Optimal Transport},
  journal = {Journal of Machine Learning Research},
  year    = {2021},
  volume  = {22},
  number  = {78},
  pages   = {1-8},
}

@misc{wang2024textembeddingsweaklysupervisedcontrastive,
  title         = {Text Embeddings by Weakly-Supervised Contrastive Pre-training},
  author        = {Liang Wang and Nan Yang and Xiaolong Huang and Binxing Jiao and Linjun Yang and Daxin Jiang and Rangan Majumder and Furu Wei},
  year          = {2024},
  eprint        = {2212.03533},
  archivePrefix = {arXiv},
  primaryClass  = {cs.CL},
  url           = {https://arxiv.org/abs/2212.03533}
}

@misc{awasthy2025graniteembeddingmodels,
  title         = {{Granite} Embedding Models},
  author        = {Parul Awasthy and Aashka Trivedi and Yulong Li and Mihaela Bornea and David Cox and Abraham Daniels and Martin Franz and Gabe Goodhart and Bhavani Iyer and Vishwajeet Kumar and Luis Lastras and Scott McCarley and Rudra Murthy and {Vignesh P} and Sara Rosenthal and Salim Roukos and Jaydeep Sen and Sukriti Sharma and Avirup Sil and Kate Soule and Arafat Sultan and Radu Florian},
  year          = {2025},
  archivePrefix = {arXiv},
  url           = {https://arxiv.org/abs/2502.20204}
}

@inproceedings{ni-etal-2022-large,
  title     = {Large Dual Encoders Are Generalizable Retrievers},
  author    = {Ni, Jianmo and Qu, Chen and Lu, Jing and Dai, Zhuyun and Hernandez Abrego, Gustavo and Ma, Ji and Zhao, Vincent and Luan, Yi and Hall, Keith and Chang, Ming-Wei and Yang, Yinfei},
  editor    = {Goldberg, Yoav and Kozareva, Zornitsa and Zhang, Yue},
  booktitle = {Proceedings of the 2022 Conference on Empirical Methods in Natural Language Processing},
  month     = dec,
  year      = {2022},
  pages     = {9844--9855},
  doi       = {10.18653/v1/2022.emnlp-main.669},
}
}


\newpage
\appendix
\setcounter{figure}{0}
\setcounter{table}{0}
\setcounter{equation}{0}
\renewcommand{\thefigure}{S\arabic{figure}}
\renewcommand{\thetable}{S\arabic{table}}
\renewcommand{\theequation}{S\arabic{equation}}

\begin{center}
\Large{\textbf{Appendix}}
\end{center}

\section{fMRI preprocessing details}\label{sec:fmri-processing}

For all main analyses, we use the Natural Scenes Dataset (NSD)~\cite{Allen2022} volumetric 1\,mm GLM beta estimates\cite{GLMSingle2022}, version 3, as provided by the dataset authors\footnote{\url{https://registry.opendata.aws/nsd/} (accessed May 2026).}. For each subject, we extract voxels from the \texttt{nsdgeneral} ROI, a reliability-based mask that includes visually responsive voxels across visual and high-level associative cortex. This yields approximately 85,000 voxels per subject. The resulting responses are organized into a trial-by-voxel matrix $X \in \mathbb{R}^{n_{\mathrm{trials}} \times v_s}$ for each subject $s$.
 
Before fitting the encoder, we apply a simple normalization procedure to reduce the effect of extreme beta values. For each subject, voxel values are clipped at the 0.05th and 99.95th percentiles and stored as the input response matrix for subsequent preprocessing. We additionally remove acquisition-related confounds by fitting a linear model of session/run structure on the training trials and subtracting the predicted confound component from all responses. All train/test splits are defined at the stimulus level before fitting any preprocessing transformations.

\subsection{Confound removal}
NSD data were acquired over multiple scanning sessions, with each subject completing up to 40 sessions and multiple runs per session. We observed that trial-by-trial correlation matrices contained visible block structure aligned with run/session order, indicating acquisition-related components in the raw fMRI responses (Fig.~S1A--B). To reduce this structure, we residualized each subject's voxel responses with respect to run/session confounds.

Let $X \in \mathbb{R}^{n_{\mathrm{trials}} \times v_s}$ denote the preprocessed beta matrix and let $C \in \mathbb{R}^{n_{\mathrm{trials}} \times c}$ denote a confound design matrix encoding the run/session membership of each trial. We fit a ridge-regularized linear confound model using only training trials:
\[
B^* = \arg\min_B \|X_{\mathrm{train}} - C_{\mathrm{train}}B\|_F^2
+ \lambda_{\mathrm{conf}}\|B\|_F^2 .
\]
The fitted confound coefficients are then applied to all trials, and residualized responses are computed as
\[
X_{\mathrm{res}} = X - CB^* .
\]
Thus, nuisance parameters are estimated exclusively from the training set and then applied to both training and held-out trials, avoiding leakage from the evaluation responses into the fitted residualization model.

\begin{figure}[!b]
    \centering
    \includegraphics[width=\linewidth]{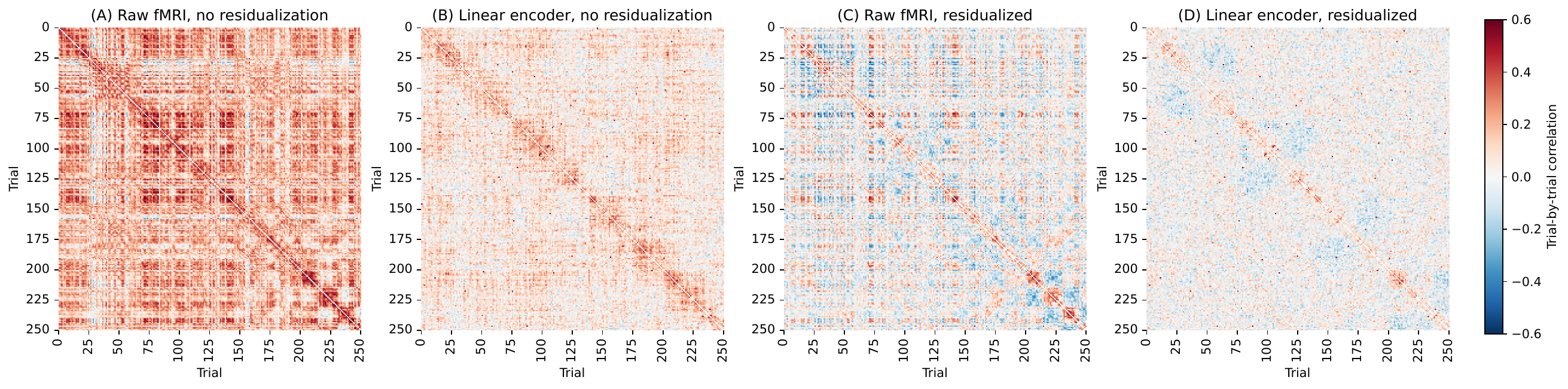}
    \caption{\textbf{Residualization mitigates acquisition-related trial structure.}
    Trial-by-trial correlation matrices for the first 250 trials from the first session of subject S1, shown for raw fMRI responses and linear encoder embeddings before and after residualizing run/session confounds. Trials are ordered by acquisition order; therefore, block-like off-diagonal structure reflects correlations due to scanning confounds rather than stimulus identity. Raw fMRI responses show pronounced run/session structure before residualization, which is partially mitigated after confound regression. The linear encoder further attenuates acquisition-related correlations. Diagonal entries are omitted for visualization.}
    \label{fig:residual_correlations}
\end{figure}

Residualization partially mitigates the run/session block structure visible in trial-wise correlation matrices (Fig.~\ref{fig:residual_correlations}). All experiments and baselines in the main text use these residualized fMRI responses as input.

\subsection{Brain region selection}
For the main analyses, we extract neural responses from the NSDGeneral ROI provided by the NSD authors~\cite{Allen2022}. This mask includes visually responsive voxels across visual and high-level associative cortex selected based on stimulus-related reliability. We use the volumetric preprocessing for the main experiments.

Table~\ref{tab:roi_preprocessing_ablation} evaluates the effect of cortical region and preprocessing space using the same encoder configuration as in the main analyses (see Section~\ref{sec:ablations}). Encoder quality is measured with within-subject retrieval across repeated presentations. Performance improves as the cortical mask expands from V1 to V1--V4 and NSDGeneral, with NSDGeneral and Visual Streams yielding the best retrieval and RSA. Volumetric and surface-based NSDGeneral preprocessing give similar performance, so we use volumetric NSDGeneral responses for all main reported analyses.

\begin{table}[ht]
\centering
\caption{\textbf{Effect of brain region and preprocessing on encoder performance.}
Within-subject retrieval across repeated presentations for different cortical regions (V1, V1--V4, NSDGeneral, Kastner, and visual streams; combining both hemispheres) and preprocessing spaces (volumetric and surface-based). Results are averaged across subjects and repetition pairs. Lower is better for Mean Rank (chance $=258$); higher is better for R@1 (chance=0.002) and RSA.}
\label{tab:roi_preprocessing_ablation}
\resizebox{\linewidth}{!}{%
    \begin{tabular}{llr|ccc}
\toprule
\textbf{Brain region} & \textbf{Preprocessing} & \textbf{Region size} 
& \textbf{Mean Rank} $\downarrow$ & \textbf{R@1} $\uparrow$ & \textbf{RSA} $\uparrow$ \\
\midrule
V1 \cite{Glasser2016} & Volumetric 1mm & 19,820 voxels & 23.42 & 0.54 & 0.37 \\
V1--V4 \cite{Glasser2016} & Volumetric 1mm & 51,197 voxels & 8.20 & 0.74 & 0.47 \\
Kastner2015 \cite{Kastner2015} & Volumetric 1mm & 56,743 voxels & 6.05 & 0.80 & 0.51 \\
NSDGeneral \cite{Allen2022} & Volumetric 1mm & 84,848 voxels & 5.28 & 0.82 & 0.53 \\
Streams \cite{Allen2022} & Volumetric 1mm & 152,370 voxels & 5.18 & 0.82 & 0.53 \\
\midrule
NSDGeneral \cite{Allen2022} & Native surface & 68,464 vertices & 5.59 & 0.81 & 0.52 \\
\bottomrule
\end{tabular}
}
\end{table}

\section{Encoder details and ablations}\label{sec:ablations}

\subsection{Encoder implementation details}

All subject encoders use the same hyperparameters. For the linear stage, we use $d_{\mathrm{PCA}}=768$ PCA components and an embedding dimensionality of $d=128$. The nonlinear residual refinement is a one-hidden-layer MLP with hidden size $d_{\mathrm{hidden}}=768$. The residual scaling parameter $\alpha$ is learned jointly with the MLP parameters and converges to $\alpha \approx 0.4$ across subjects.

The MLP is optimized with Adam for 2000 training steps using the combined objective described in the main text, with $\lambda_{\mathrm{NCE}}=1$ and $\lambda_{\mathrm{pull}}=0.5$. Hyperparameters were selected by within-subject retrieval across repeated presentations, balancing Mean Rank and R@1. Unless stated otherwise, all experiments use these settings, yielding 128-dimensional subject embeddings.

\begin{table}[b]
\centering
\caption{\textbf{Encoder component ablation}.
Within-subject retrieval across repeated presentations after removing components of the encoder. Results are averaged across subjects and repetition pairs. Lower is better for Mean Rank; higher is better for R@1 and RSA.}
\label{tab:encoder_component_ablation}
\resizebox{\linewidth}{!}{%
    \begin{tabular}{lccccccc}
\toprule
\textbf{Configuration} 
& \textbf{Reliability} & \textbf{PCA} & \textbf{MCCA} & \textbf{Nonlinear}
& \textbf{Mean Rank} $\downarrow$ & \textbf{R@1} $\uparrow$ & \textbf{RSA} $\uparrow$ \\
\midrule
Reliability + PCA & $\checkmark$ & $\checkmark$ & -- & -- & 116.52 & 0.09 & 0.13 \\
PCA + MCCA & -- & $\checkmark$ & $\checkmark$ & -- & 77.31 & 0.20 & 0.27 \\
Reliability + PCA + MCCA & $\checkmark$ & $\checkmark$ & $\checkmark$ & -- & 11.73 & 0.73 & 0.23 \\
\midrule
PCA + nonlinear & -- & $\checkmark$ & -- & $\checkmark$ & 14.05 & 0.60 & 0.55 \\
PCA + MCCA + nonlinear & -- & $\checkmark$ & $\checkmark$ & $\checkmark$ & 14.05 & 0.60 & 0.54 \\
Reliability + PCA + nonlinear & $\checkmark$ & $\checkmark$ & -- & $\checkmark$ & 21.89 & 0.43 & 0.51 \\
Full encoder & $\checkmark$ & $\checkmark$ & $\checkmark$ & $\checkmark$ & \textbf{5.28} & \textbf{0.82} & \textbf{0.53} \\
\bottomrule
\end{tabular}
}
\end{table}

\subsection{Encoder component ablation.}
We ablated the main components of the encoder by removing reliability weighting, MCCA, and the nonlinear residual refinement. Table~\ref{tab:encoder_component_ablation} shows that the full encoder gives the best retrieval performance. The nonlinear refinement contributes strongly to RSA and retrieval, while the full combination of reliability weighting, PCA, MCCA, and nonlinear refinement yields the best Mean Rank and R@1.

\subsection{Encoder dimensionality sensitivity}
\begin{figure}[h]
    \centering
    \includegraphics[width=\linewidth]{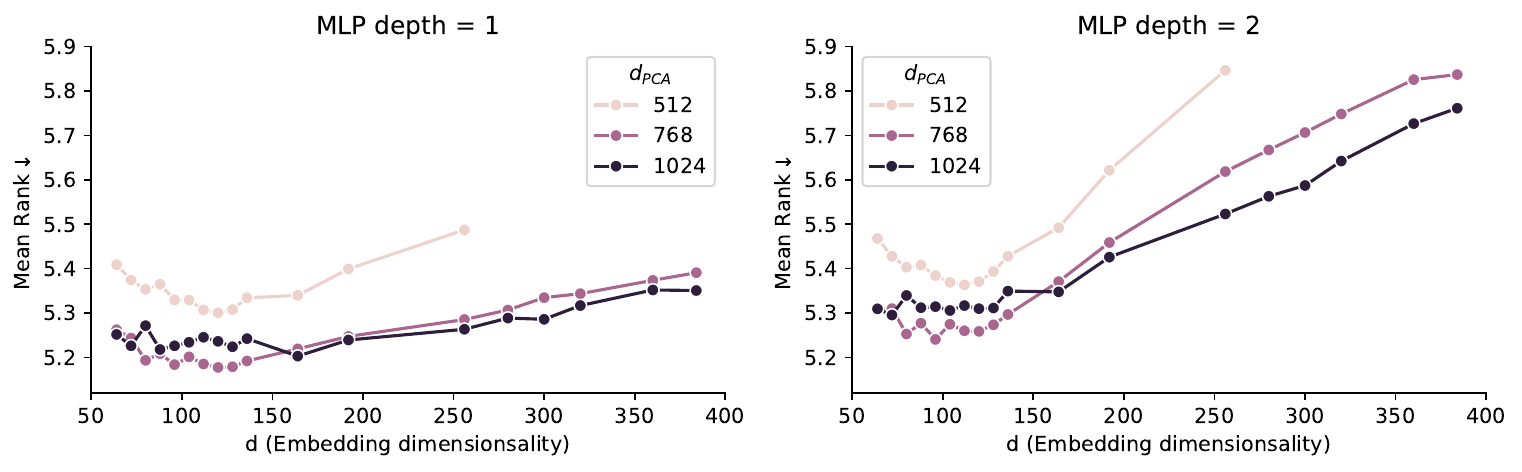}
     \caption{\textbf{Encoder dimensionality sensitivity.}
    Mean Rank across encoder dimensionalities. Lower is better. Each point reports the best value across $d_{hidden}$ sizes. Mean Rank averaged across subjects.}
    \label{fig:encoder_dimensionality_sensitivity}
\end{figure}

We evaluated within-subject retrieval across a hyperparameter grid varying PCA dimensionality $d_{\mathrm{PCA}} \in \{512,768,1024,1280\}$, embedding dimensionality $d \in [64,384]$, hidden size $d_{\mathrm{hidden}} \in \{128,256,512,768,1024\}$, and MLP depth $\in \{1,2,3\}$. Fig.~\ref{fig:encoder_dimensionality_sensitivity} summarizes the search by plotting Mean Rank as a function of $d$, with separate curves for $d_{\mathrm{PCA}}$ and panels for MLP depth. Since $d_{\mathrm{hidden}}$ is not shown, each point reports the best Mean Rank across hidden sizes for that combination of $d$, $d_{\mathrm{PCA}}$, and depth. Overall, a single hidden layer performs best, and the strongest stable configurations use $d_{\mathrm{PCA}}=768$ or $1024$ with embedding dimensionality around $d=100$--$150$.

\section{Unsupervised selection}\label{sec:unsupervised-selection}

The pairwise translation procedure (Section \ref{subsec:pairwise}) is sensitive to its initialization. The pseudo-correspondences obtained through centroid matching depend on the K-means initialization\cite{Dar2026}. Different random seeds produce different initial rotations and converge to different local solutions, some of which correspond to poor alignments.

To address this sensitivity without using paired samples or evaluation labels, we run the pairwise translation procedure with \(C\) random seeds. For every ordered subject pair \((s,t)\), this produces a set of candidate rotations

\begin{equation}
\mathcal{R}_{s\rightarrow t}
=
\left\{
R_{s\rightarrow t}^{(c)}
\right\}_{c=1}^{C},
\qquad
R_{s\rightarrow t}^{(c)}\in O(d).
\end{equation}

We select one candidate per subject pair using cycle consistency. For three distinct subjects \(s,t,p\), a coherent set of rotations should satisfy

\begin{equation}
R_{s\rightarrow t}R_{t\rightarrow p}
\approx R_{s\rightarrow p}.
\end{equation}

We measure agreement between two orthogonal transformations \(A\) and \(B\) using their normalized Frobenius inner product,

\begin{equation}
s_{\mathrm{rot}}(A,B)
=
\frac{1}{d}\operatorname{tr}(A^\top B).
\end{equation}

This score is \(1\) for identical transformations and is expected to be close to \(0\) for unrelated rotations. Let \(c_{st}\in\{1,\ldots,C\}\) denote the candidate selected for pair \((s,t)\), and let \(\mathbf{c}=\{c_{st}\}_{s\neq t}\) denote a complete assignment of candidates to all ordered subject pairs. We define the global cycle-consistency objective as

\begin{equation}
J(\mathbf{c})
=
\frac{1}{|\mathcal{T}|}
\sum_{(s,t,p)\in\mathcal{T}}
s_{\mathrm{rot}}\left(
R_{s\rightarrow t}^{(c_{st})}
R_{t\rightarrow p}^{(c_{tp})},
R_{s\rightarrow p}^{(c_{sp})}
\right),
\end{equation}

where \(\mathcal{T}\) contains all ordered triplets of distinct subjects.

We optimize this discrete objective using coordinate ascent \cite{Wright2015}:

\begin{enumerate}
    \item Randomly select one candidate for every ordered subject pair.
    \item Visit each ordered pair and select the candidate that maximizes \(J\), while keeping all other selections fixed.
    \item Repeat these sweeps until no selection changes.
    \item Repeat steps 1–3 from \(M\) random initializations and retain the selection with the largest value of \(J\).
\end{enumerate}

The criterion depends exclusively on the candidate rotation matrices and does not use training embeddings, evaluation embeddings, image identities, or retrieval performance. The resulting rotations,
$
\left\{
R_{s\rightarrow t}^{(c_{st}^{\star})}
\right\}_{s\neq t},
$
are used as the pairwise translations in all subsequent analyses.

\subsection{Unsupervised selection stability}
\label{sec:selection_stability}

To quantify the stability of the unsupervised selection algorithm, we first ran the pairwise translation procedure with \(C=10\) random seeds, obtaining 10 candidate rotations for each ordered subject pair. Keeping these candidate sets fixed, we ran the coordinate-ascent selection algorithm 100 times with different global random seeds, using \(M=100\) random restarts in each run. We evaluated Mean Rank and Recall@1 for the selected pairwise rotations, both before and after the spectral synchronization described in Section~3.3. The evaluation metrics were used only to characterize the resulting solutions and were not accessed by the selection algorithm. The same selection of 56 directed rotations was recovered in 63 of the 100 runs. All main results use this most frequently recovered selection.

\begin{table}[hb]
    \centering
    \caption{
        Stability of the unsupervised rotation-selection algorithm over 100 runs. Values in the first row report the mean and standard deviation across runs.
    }
    \label{tab:selection_stability}
    \resizebox{\linewidth}{!}{
    \begin{tabular}{lcccc}
        \toprule
        Statistic
        & \begin{tabular}[c]{@{}c@{}}Mean Rank\\(pairwise) \(\downarrow\)\end{tabular}
        & \begin{tabular}[c]{@{}c@{}}Recall@1\\(pairwise) \(\uparrow\)\end{tabular}
        & \begin{tabular}[c]{@{}c@{}}Mean Rank\\(synchronized) \(\downarrow\)\end{tabular}
        & \begin{tabular}[c]{@{}c@{}}Recall@1\\(synchronized) \(\uparrow\)\end{tabular} \\
        \midrule
        Mean \(\pm\) std
        & \textbf{2.7046 \(\pm\) 0.0245}
        & \textbf{0.7800 \(\pm\) 0.0003}
        & \textbf{2.1038 \(\pm\) 0.0064}
        & \textbf{0.8213 \(\pm\) 0.0004} \\
        Median
        & 2.7002
        & 0.7801
        & 2.1080
        & 0.8215 \\
        Most frequent solution (63/100)
        & 2.7002
        & 0.7801
        & 2.1080
        & 0.8215 \\
        Minimum
        & 2.6694
        & 0.7794
        & 2.0799
        & 0.8203 \\
        Maximum
        & 2.7725
        & 0.7804
        & 2.1080
        & 0.8215 \\
        \bottomrule
    \end{tabular}
    
    }
\end{table}

\section{Experimental details}\label{sec:details}

This section reports additional details to reproduce the evaluation and main results.

\subsection{Within-subject encoder baseline details}

To assess whether encoder improvements were consistent across subjects, Table~\ref{tab:within_subject_baselines_subjects} reports Mean Rank for each within-subject baseline separately for all subjects (extending Table \ref{tab:within_subject_main}).

\begin{table}[h]
\centering
\caption{\textbf{Per-subject within-subject retrieval across encoder baselines.}
Mean Rank for within-subject retrieval across repeated presentations, reported separately for each subject and averaged across repetition pairs. Lower is better; chance Mean Rank is $258$.}
\label{tab:within_subject_baselines_subjects}
\resizebox{\linewidth}{!}{%
    \begin{tabular}{l|cccccccc|c}
\toprule
\textbf{Method}
& \textbf{S1} & \textbf{S2} & \textbf{S3} & \textbf{S4}
& \textbf{S5} & \textbf{S6} & \textbf{S7} & \textbf{S8}
& \textbf{Avg.} \\
\midrule
Preprocessed fMRI (GLMsingle~\cite{GLMSingle2022})
& 114.48 & 122.62 & 155.37 & 154.53 & 118.40 & 150.04 & 155.55 & 182.19 & 144.15 \\
\midrule
PCA
& 54.28 & 35.95 & 98.65 & 83.27 & 68.03 & 86.60 & 93.09 & 104.93 & 78.10 \\
PCA + SplitAE~\cite{SplitAE}
& 60.99 & 58.23 & 104.29 & 101.37 & 65.04 & 105.20 & 113.08 & 142.11 & 93.78 \\
PCA + DeepCCA~\cite{DeepCCA}
& 1.94 & 4.72 & 31.67 & 22.27 & 15.75 & 22.20 & 30.93 & 56.18 & 23.21 \\
PCA + Multiview CCA~\cite{RegMCCA2011}
& 1.40 & 2.79 & 27.95 & 19.78 & 9.57 & 17.71 & 19.72 & 54.92 & 19.23 \\
\midrule
DINOv2 ViT-S/14~\cite{dinov2} + Ridge
& 11.43 & 16.42 & 28.99 & 26.76 & 13.12 & 20.75 & 29.60 & 51.52 & 24.82 \\
CLIP ViT-B/16~\cite{Radford2021} + Ridge
& 12.30 & 16.88 & 27.1 & 24.86 & 11.71 & 18.1 & 28.25 & 46.66 & 23.23 \\
ViT AugReg-L/16~\cite{steiner2022how} + Ridge
& 7.93 & 12.9 & 24.51 & 22.98 & 10.22 & 15.4 & 24.37 & 44.12 & 20.30 \\
all-MiniLM-L6-v2~\cite{SentenceBert} + Ridge
& 18.67 & 24.80 & 35.63 & 31.97 & 18.24 & 25.45 & 34.00 & 55.00 & 30.47 \\
\midrule
Ours (linear)
& 1.09	& 1.20 & 20.55 & 13.4 & 4.54 & 8.42 & 15.38 & 29.26 & 11.73 \\
Ours (full)
& \textbf{1.03} & \textbf{1.12} & \textbf{11.25} & \textbf{4.69} & \textbf{1.77} & \textbf{3.09} & \textbf{7.60} & \textbf{11.67} & \textbf{5.28} \\
\bottomrule
\end{tabular}
}
\end{table}

We evaluated model-guided baselines by fitting a linear ridge regression from PCA-reduced fMRI responses (768 components) to intermediate model representations, and then measuring within-subject retrieval across repeated presentations. Vision embeddings were extracted from the last layer of pretrained models applied to the NSD images using the TIMM library\cite{rw2019timm}, while language embeddings were extracted from the COCO captions associated with each image using sentence-transformers\cite{SentenceBert}. Language models follow those used in~\cite{Dar2026}, and vision models were selected from~\cite{MarcosManchon2025}. A complete list of model identifiers is provided as a Hugging Face collection\footnote{https://huggingface.co/collections/pablomm/platonic-representations-in-the-human-brain}. The main text reports representative models from the main model families in Table~\ref{tab:within_subject_main}. Table~\ref{tab:model_guided_baselines_subjects} provides the detailed per-subject evaluation for the larger set of model representations tested.

\begin{table}[h]
\centering
\caption{\textbf{Per-subject model-guided ridge baselines.}
Mean Rank for within-subject retrieval across repeated presentations using ridge mappings from fMRI responses to pretrained vision- and language-model embeddings. Results are reported separately for each subject and averaged across subjects. Lower is better; chance Mean Rank is $258$.}
\label{tab:model_guided_baselines_subjects}
\resizebox{\linewidth}{!}{%
    \begin{tabular}{l|cccccccc|c}
\toprule
\textbf{Model}
& \textbf{S1} & \textbf{S2} & \textbf{S3} & \textbf{S4}
& \textbf{S5} & \textbf{S6} & \textbf{S7} & \textbf{S8}
& \textbf{Avg.} \\
\midrule
\multicolumn{10}{l}{\textit{Vision models}} \\
\midrule
DINOv2 ViT-S/14 \cite{dinov2} & 11.43 & 16.42 & 28.99 & 26.76 & 13.12 & 20.75 & 29.60 & 51.52 & 24.82 \\
DINOv2 ViT-B/14  & 19.11 & 24.55 & 35.68 & 38.05 & 20.33 & 32.07 & 40.18 & 65.86 & 34.48 \\
DINOv2 ViT-L/14 & 18.57 & 24.35 & 38.20 & 39.97 & 19.70 & 32.75 & 42.43 & 73.26 & 36.15 \\
DINOv2 ViT-G/14 & 26.51 & 30.49 & 44.01 & 46.19 & 24.07 & 39.55 & 49.98 & 77.67 & 42.31 \\
\midrule
AugReg ViT-T/16 \cite{steiner2022how} & 14.33 & 17.77 & 29.29 & 27.42 & 14.32 & 20.53 & 30.00 & 48.51 & 25.27 \\
AugReg ViT-S/16 & 11.87 & 17.66 & 27.84 & 26.57 & 12.26 & 19.31 & 29.49 & 48.57 & 24.20 \\
AugReg ViT-B/16 & 8.63 & 13.95 & 25.15 & 23.07 & 10.99 & 16.27 & 25.25 & 43.62 & 20.87 \\
AugReg ViT-L/16 & \textbf{7.93} & \textbf{12.90} & \textbf{24.51} & \textbf{22.98} & \textbf{10.22} & \textbf{15.40} & \textbf{24.37} & \textbf{44.12} & \textbf{20.30} \\
\midrule
MAE ViT-B/16 \cite{mae} & 18.67 & 25.38 & 40.77 & 35.04 & 20.81 & 30.20 & 41.84 & 59.57 & 34.03 \\
MAE ViT-L/16 & 12.63 & 17.72 & 32.39 & 27.66 & 14.20 & 22.66 & 33.31 & 50.08 & 26.33 \\
MAE ViT-H/14 & 13.23 & 18.17 & 32.98 & 27.89 & 14.12 & 23.25 & 34.80 & 49.73 & 26.77 \\
\midrule
CLIP ViT-B/16 FT \cite{Radford2021} & 17.97 & 23.00 & 34.80 & 34.59 & 17.65 & 26.87 & 36.85 & 61.08 & 31.60 \\
CLIP ViT-L/14 FT & 14.08 & 20.36 & 31.19 & 30.23 & 14.18 & 23.15 & 33.49 & 56.79 & 27.93 \\
CLIP ViT-H/14 FT & 16.70 & 23.27 & 35.24 & 33.82 & 16.89 & 26.59 & 38.52 & 62.65 & 31.71 \\
\midrule
CLIP ViT-B/16 \cite{Radford2021} & 12.30 & 16.88 & 27.10 & 24.86 & 11.71 & 18.10 & 28.25 & 46.66 & 23.23 \\
CLIP ViT-L/14 & 12.78 & 17.99 & 28.57 & 25.85 & 12.36 & 19.14 & 29.21 & 48.99 & 24.36 \\
CLIP ViT-H/14 & 10.86 & 15.74 & 26.51 & 23.78 & 11.36 & 17.54 & 26.88 & 46.87 & 22.44 \\
\midrule
\multicolumn{10}{l}{\textit{Language models}} \\
\midrule
E5-small \cite{wang2024textembeddingsweaklysupervisedcontrastive} & 24.19 & 30.91 & 42.11 & 38.52 & 21.75 & 31.60 & 42.15 & 65.67 & 37.11 \\
E5 & 22.92 & 30.18 & 41.78 & 38.04 & 20.71 & 31.02 & 41.41 & 65.26 & 36.41\\
Stella & 22.34 & 29.00 & 39.95 & 36.62 & 19.86 & 29.19 & 39.28 & 63.24 & 34.93 \\
Granite Embedding-small \cite{awasthy2025graniteembeddingmodels} & 22.50 & 27.69 & 38.77 & 34.96 & 19.04 & 29.05 & 38.55 & 60.79 & 33.92 \\
Granite Embedding & 21.78 & 27.77 & 38.80 & 34.80 & 19.38 & 28.58 & 37.63 & 61.37 & 33.76\\
GTR \cite{ni-etal-2022-large} & 19.49 & 24.90 & 36.21 & 33.78 & 16.95 & 26.84 & 35.83 & 58.75 & 31.59\\
all-MiniLM-L6-v2 & 18.67 & 24.80 & 35.63 & 31.97 & 18.24 & 25.45 & 34.00 & 55.00 & 30.47 \\
\bottomrule
\end{tabular}
}
\end{table}

\subsection{Cross-subject alignment baselines}
\label{sec:cross_subject_baselines}

All baselines for the cross-subject alignment task use repetition-averaged representations and
are evaluated on the same 515 shared images.
Results are averaged across the 56 directed subject pairs.

\textbf{PCA without alignment.}
We independently fit 768-component PCA representations
for each subject and compare them directly, with components
ordered by decreasing explained variance.
No cross-subject mapping is fitted.

\textbf{Optimal transport.}
We use entropic Gromov--Wasserstein matching from the
Python Optimal Transport (POT) library~\cite{flamary2021pot}.
The method computes within-subject distance matrices
and estimates a transport plan that matches their
relational structure across subjects.
We apply it to the complete evaluation sets of PCA
representations and learned embeddings, without providing
the true image correspondences.

\textbf{Model-guided ridge regression.}
For each subject, ridge regression maps neural responses
to model features of its own training images.
Cross-subject retrieval is then evaluated in the common
model feature space. These baselines use unpaired training
images but require external image or caption representations,
extracted as described above.

\textbf{Supervised alignment protocol.}
The following methods use our fixed 128-dimensional
embeddings and up to 485 additional shared training images
per subject pair. The 515 evaluation images are excluded
from fitting these mappings, and the encoders remain
trained only on non-overlapping images.

\textbf{Orthogonal Procrustes (supervised).}
We fit an orthogonal mapping that minimizes squared error
between paired source and target embeddings, using singular
value decomposition. The mapping is then applied to source
evaluation embeddings.

\textbf{Cross-validated ridge regression (supervised).}
We fit an affine source-to-target mapping with an $\ell_2$
penalty on its weights. Regularization is selected from
$\{10^{-4},10^{-3},10^{-2},10^{-1},0.2,0.5,1\}$
by five-fold cross-validation minimizing mean squared error,
with shuffled folds and random seed 42.
The selected model is refitted on all paired training images.

\textbf{Canonical correlation analysis (CCA) (supervised).}
We use regularized CCA from \texttt{mvlearn}, with
128 components and regularization parameter $0.3$.
Separate subject projections are fitted to paired training
embeddings, and retrieval is evaluated between their
projected evaluation embeddings.

\textbf{Linear regression.}
We fit an unregularized affine source-to-target mapping
by least squares and apply it to source evaluation embeddings.

\subsection{Compute requirements}
All experiments were run on a single workstation with an NVIDIA A40 GPU (48 GB) and an AMD Ryzen 9 CPU. The experiments are not computationally intensive: training the linear encoder for one subject takes approximately 2 minutes on CPU, training the nonlinear residual MLP takes approximately 1 minute on GPU, and estimating one pairwise brain-to-brain rotation takes approximately 5 minutes on CPU.

\section{Statistical support and robustness}
\label{sec:statistical_analysis}

We assess statistical significance and uncertainty in cross-subject
retrieval (Fig.~\ref{fig:global_alignment}) using image-identity
permutation tests, image bootstrapping, and subject-deletion
sensitivity. All analyses use the synchronized rotations defining
the shared space reported in Section~\ref{sec:shared-alignment-results},
with encoders and rotations held fixed. We evaluate 515 held-out
shared images across 56 directed pairs of eight NSD participants
and average retrieval metrics across pairs.

\begin{table}[h]
\centering
\small
\caption{\textbf{Image-level statistical support for
synchronized retrieval.}
Global metrics are averaged across 56 directed subject pairs.
Null and bootstrap summaries report mean $\pm$ standard
deviation across $10{,}000$ permutations or resamples.
Confidence intervals are percentile image-bootstrap intervals.
}
\label{tab:retrieval_image_statistics}
\resizebox{\linewidth}{!}{%
\begin{tabular}{lcccc}
\toprule
Metric & Observed & Permutation null & Bootstrap & 95\% CI \\
\midrule
R@1 $\uparrow$
& 0.8215
& $0.00192 \pm 0.00163$
& $0.8215 \pm 0.0068$
& $[0.8079, 0.8350]$ \\
Mean Rank $\downarrow$
& 2.108
& $258.011 \pm 5.191$
& $2.107 \pm 0.118$
& $[1.893, 2.357]$ \\
\bottomrule
\end{tabular}
}
\end{table}

\subsection{Image-identity permutation tests}

We tested retrieval against chance using $10{,}000$ random
permutations of image identities on one side of each comparison.
Each permutation was shared across all directed subject pairs,
and metrics were averaged across pairs.
No permutation yielded a Mean Rank as low as observed or
an R@1 as high as observed. Using the one-sided permutation
$p$-value $(k+1)/(B+1)$, where $k$ counts equally or more
extreme results, both metrics gave $p<10^{-4}$
(Table~\ref{tab:retrieval_image_statistics}).

\subsection{Subject-deletion sensitivity}

We assessed sensitivity to individual participants by excluding
each subject's evaluation pairs and averaging metrics over the
remaining 42 directed pairs, with the fitted pipeline held fixed.
We computed approximate 95\% jackknife intervals using
subject-deletion standard errors and a $t$ distribution with
seven degrees of freedom.
Mean Rank ranged from $1.909$ to $2.261$ and R@1 from
$0.8032$ to $0.8492$, showing strong retrieval after each
exclusion (Table~\ref{tab:retrieval_subject_sensitivity}).

\begin{table}[ht]
\centering
\small
\caption{\textbf{Subject-deletion sensitivity.}
Metrics exclude each subject's evaluation pairs in turn,
keeping the fitted eight-subject shared space fixed.
Intervals are approximate 95\% jackknife intervals.}
\label{tab:retrieval_subject_sensitivity}
\resizebox{0.75\linewidth}{!}{%
\begin{tabular}{lccc}
\toprule
Metric & Full result & Subject-deletion range & 95\% interval \\
\midrule
R@1 $\uparrow$
& 0.8215 & $0.8032$--$0.8492$ & $[0.7337, 0.9093]$ \\
Mean Rank $\downarrow$
& 2.108 & $1.909$--$2.261$ & $[1.342, 2.874]$ \\
\bottomrule
\end{tabular}
}
\end{table}

\subsection{Image-bootstrap confidence intervals}

To quantify uncertainty due to image sampling, we computed
$10{,}000$ bootstrap resamples of 515 query-image indices
with replacement, sharing sampled indices across subject
pairs while keeping the candidate set fixed.
The 95\% percentile intervals were $[1.893, 2.357]$ for
Mean Rank and $[0.8079, 0.8350]$ for R@1
(Table~\ref{tab:retrieval_image_statistics}).

Table~\ref{tab:pairwise_bootstrap} reports 95\% image-bootstrap
confidence intervals for each directed subject pair.
Pairwise Mean Rank ranged from $1.04$ to $4.56$, and all
intervals remained below the chance value of $258$.

\begin{table}[p]
    \centering
    \setlength{\tabcolsep}{3pt}
    \caption{\textbf{Pairwise cross-subject retrieval uncertainty.}
    Each cell reports Mean Rank above its pointwise 95\%
    image-bootstrap confidence interval.
    Lower Mean Rank is better; chance is $258$.}
    \label{tab:pairwise_bootstrap}
    \begin{tabular}{@{}lcccccccc@{}}
    \toprule
    \textbf{Subject}
    & \textbf{S1} & \textbf{S2} & \textbf{S3} & \textbf{S4}
    & \textbf{S5} & \textbf{S6} & \textbf{S7} & \textbf{S8} \\
    \midrule
    \textbf{S1}
    & ---
    & \rankci{1.04}{1.02}{1.06}
    & \rankci{1.44}{1.18}{1.81}
    & \rankci{1.68}{1.44}{1.98}
    & \rankci{4.19}{3.44}{5.02}
    & \rankci{3.39}{2.35}{4.70}
    & \rankci{1.36}{1.22}{1.52}
    & \rankci{1.93}{1.65}{2.25} \\
    \addlinespace
    \textbf{S2}
    & \rankci{1.04}{1.02}{1.06}
    & ---
    & \rankci{1.23}{1.15}{1.32}
    & \rankci{1.76}{1.48}{2.09}
    & \rankci{4.56}{3.69}{5.54}
    & \rankci{3.32}{2.52}{4.25}
    & \rankci{1.57}{1.31}{1.90}
    & \rankci{1.93}{1.54}{2.52} \\
    \addlinespace
    \textbf{S3}
    & \rankci{1.44}{1.18}{1.84}
    & \rankci{1.23}{1.14}{1.32}
    & ---
    & \rankci{1.30}{1.17}{1.44}
    & \rankci{2.31}{1.82}{2.93}
    & \rankci{2.17}{1.64}{2.81}
    & \rankci{1.35}{1.20}{1.54}
    & \rankci{1.91}{1.46}{2.60} \\
    \addlinespace
    \textbf{S4}
    & \rankci{1.72}{1.44}{2.06}
    & \rankci{1.65}{1.39}{1.96}
    & \rankci{1.36}{1.22}{1.53}
    & ---
    & \rankci{1.88}{1.60}{2.21}
    & \rankci{1.70}{1.42}{2.02}
    & \rankci{1.33}{1.20}{1.46}
    & \rankci{2.18}{1.74}{2.72} \\
    \addlinespace
    \textbf{S5}
    & \rankci{3.52}{2.92}{4.20}
    & \rankci{3.92}{3.20}{4.77}
    & \rankci{1.88}{1.52}{2.37}
    & \rankci{1.61}{1.40}{1.85}
    & ---
    & \rankci{1.52}{1.22}{1.99}
    & \rankci{2.17}{1.80}{2.58}
    & \rankci{2.86}{2.16}{3.85} \\
    \addlinespace
    \textbf{S6}
    & \rankci{3.24}{2.23}{4.48}
    & \rankci{3.24}{2.45}{4.21}
    & \rankci{2.09}{1.64}{2.64}
    & \rankci{1.75}{1.46}{2.13}
    & \rankci{1.70}{1.39}{2.11}
    & ---
    & \rankci{2.10}{1.60}{2.75}
    & \rankci{2.75}{2.18}{3.38} \\
    \addlinespace
    \textbf{S7}
    & \rankci{1.21}{1.12}{1.33}
    & \rankci{1.42}{1.21}{1.68}
    & \rankci{1.31}{1.15}{1.50}
    & \rankci{1.22}{1.13}{1.32}
    & \rankci{2.28}{1.88}{2.74}
    & \rankci{1.82}{1.43}{2.36}
    & ---
    & \rankci{2.32}{1.71}{3.11} \\
    \addlinespace
    \textbf{S8}
    & \rankci{2.11}{1.74}{2.57}
    & \rankci{2.07}{1.61}{2.77}
    & \rankci{2.09}{1.58}{2.84}
    & \rankci{2.68}{2.08}{3.39}
    & \rankci{3.47}{2.64}{4.51}
    & \rankci{3.04}{2.40}{3.74}
    & \rankci{2.72}{2.05}{3.59}
    & --- \\
    \bottomrule
\end{tabular}

\end{table}

\clearpage

\section{Qualitative Examples}
\label{sec:qualitative_examples}
\begin{figure}[h]
    \centering
    \includegraphics[width=\linewidth]{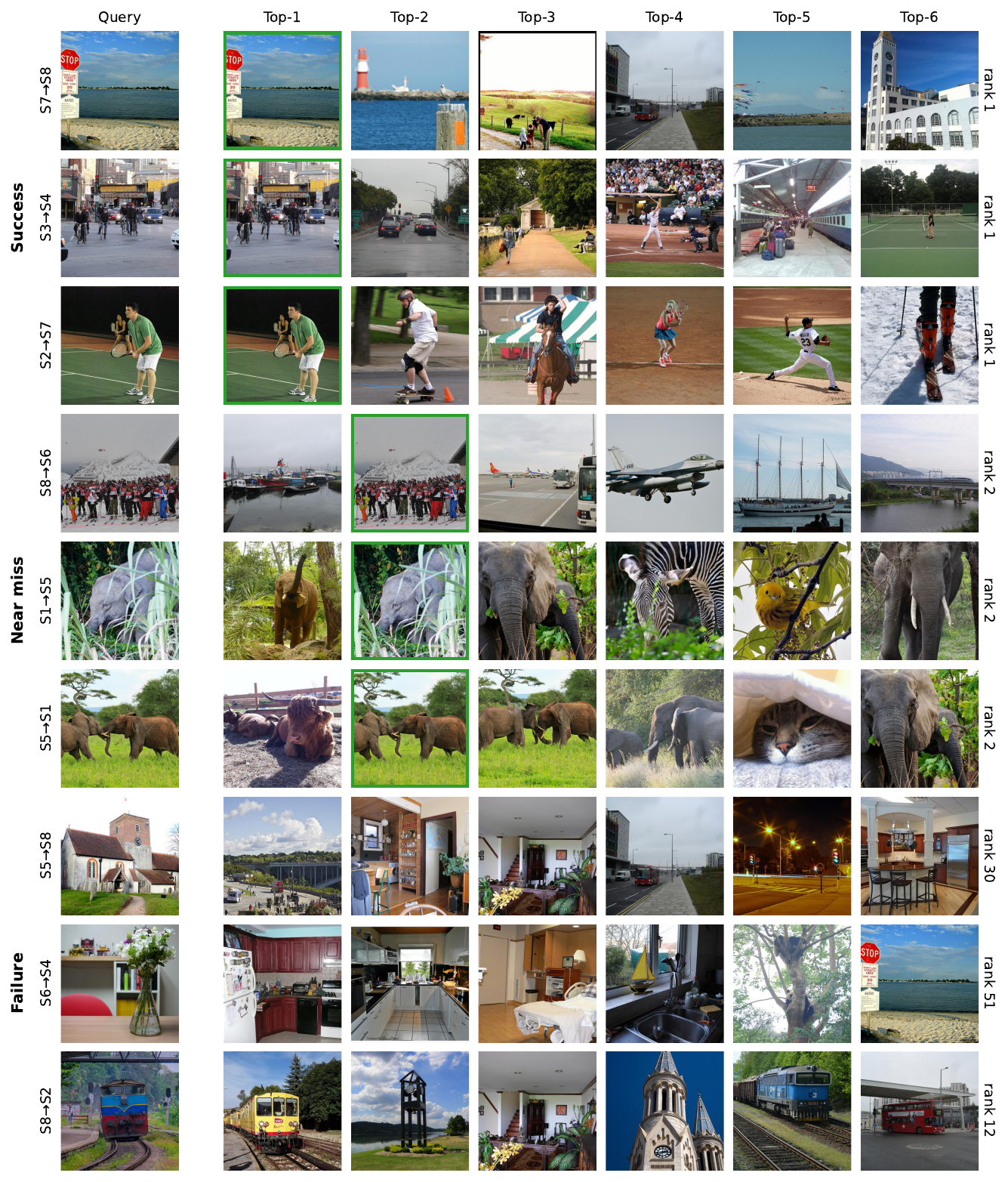}
    \caption{\textbf{Cross-subject retrieval examples in the shared space.}
Each row shows a query image and the top six images retrieved
from 515 candidates for the indicated source$\rightarrow$target
subject pair. The correct image is outlined in green when shown;
its rank appears on the right.
Examples were sampled uniformly within three strata across
all 56 directed pairs using a fixed random seed:
successes (rank 1; 82.2\%), near misses (ranks 2--3; 10.4\%),
and failures (rank $>10$; 2.6\%).}
    \label{fig:qual_success_failure}
\end{figure}

\end{document}